# The 2020 Updated Roadmaps for the US Magnet Development Program

***Compiled by***

Soren Prestemon, Kathleen Amm, Lance Cooley, Steve Gourlay, David Larbalestier, George Velev, Alexander Zlobin

&

***With Major Contributions from***
Technical Leads and Collaborators within the US MDP

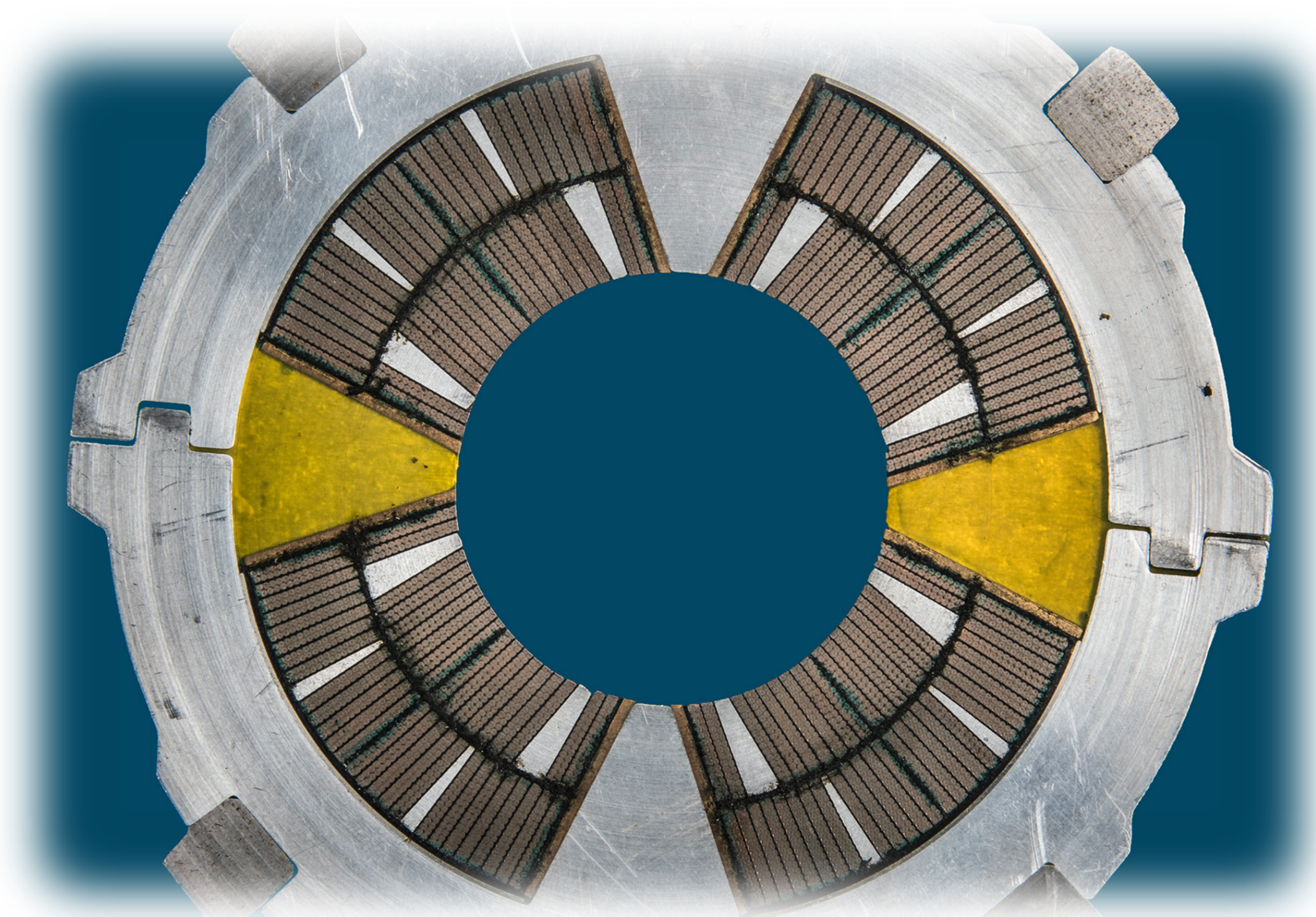

## Table of Contents

# Executive Summary

The US Magnet Development Program brings together teams from the leading US accelerator magnet research programs to develop the next generation of magnet technology for future collider applications. Sponsored by the DOE Office of High Energy Physics, the program strives to maintain and strengthen US Leadership in the field, while nurturing cross-cutting activities from other programs to further strengthen the research and its impact to the DOE Office of Science.

The US MDP was initiated in 2016, and there has been significant progress on the original program roadmaps and major advancements in magnet science. These advances, together with the addition of Brookhaven National Laboratory to the MDP in 2019 and the onset of the next Snowmass community planning effort for High Energy Physics in early 2020, motivate a review and renewal of the program roadmaps.

The updated roadmaps are the result of significant discussion and planning with the MDP research staff, along with guidance from our Technical Advisory Committee, outreach to colleagues from the International Community, and feedback from DOE-OHEP. We note that although these roadmaps are aggressive and assume a growth in program funding, they are by no means all-inclusive of research that the MDP team believe are worthy of pursuit for High Energy Physics – many research avenues are not addressed in the current program. To partially address this we have included elements in the program that enable us to develop plans so those research areas can be rapidly developed, should funding become available. We are also keenly aware of ongoing research in the international community, and our roadmaps take those efforts into consideration.

The major themes for the updated roadmaps include:

- Explore the potential for stress-managed structures to enable high-field accelerator magnets, i.e. structures that mitigate degradation to strain-sensitive $Nb_3Sn$ and HTS superconductors in high-field environments;
- Explore the potential for hybrid HTS/LTS magnets for cost-effective high field accelerator magnets that exceed the field strengths achievable with LTS materials;
- Advance magnet science through the rapid development and deployment of unique diagnostics and modeling tools to inform and accelerate magnet design improvements;
- Perform design studies on high field accelerator magnet concepts to inform DOE-OHEP on further promising avenues for magnet development;
- Advance superconductors through enhanced performance, improved production quality, and reduction in cost - all critical elements for future collider applications.

These themes are consistent with the original US MDP goals and leverage the major advances the US MDP has achieved to date in advancing superconductors, developing core HTS magnet technologies, and demonstrating record $Nb_3Sn$ accelerator magnet performance. Together these themes form the foundation for a program that will maintain US leadership in developing advance accelerator magnet technology for the years to come.

# Introduction

The US Magnet Development Program is a DOE Office of High Energy Physics (DOE-OHEP) sponsored program founded in 2016 to develop magnet technology for future collider applications. The 2016 US Magnet Development Program Plan[1] provided the motivation, strategy, and goals for the original program, along with roadmaps for the key program elements and associated technical milestones. The document furthermore identified the initial collaboration members (LBNL, FNAL, and ASC/NHMFL) and the management structure defined to clarify roles and responsibilities and to provide requisite program oversight.

The US MDP is now about four years old and is fully functioning as an integrated effort. It encompasses an energetic group of scientists, engineers, and technical staff from the collaborating institutions (*Appendix II: The MDP Community*), contributing to advance our program goals. The program has a well-established and very active Technical Advisory Committee (TAC) providing technical guidance to the program, and a Steering Council composed of DOE Representatives and Laboratory Management to provide support and oversight. Furthermore, the US MDP has grown, with the addition of Brookhaven National Laboratory to the program in 2019, formalized in a renewed Memorandum of Agreement between the collaborating members. The organization and management structure for the US MDP is outlined in *Appendix III: Management Structure.*

As we write now, we can say that the MDP is a dynamic program, with strong first results to point to and an enhanced and talented team of scientists and engineers eager to pursue magnet research. With the encouragement of DOE-OHEP and our Technical Advisory Committee, we have refreshed our Program Plan with updated roadmaps, a fine-tuning of our program focus, and refreshed strategic elements aimed at future opportunities.

# Program Overview

The mission of the US Magnet Development Program (MDP) is to perform research on advanced superconducting accelerator magnet technology for future HEP colliders. To that end, we have identified a number of driving questions and overarching goals that provide direction and focus to the program. The program is structured to address these goals and driving questions with newly-developed roadmaps for each of the major technical elements to guide the research and identify program milestones.

## Program vision and overarching goals

As a National Program composed of multiple DOE Laboratories and University members, the US MDP aspires to provide broad leadership in accelerator magnet technology. The vision of the US MDP is to:

[1] https://www2.lbl.gov/LBL-Programs/atap/MagnetDevelopmentProgramPlan.pdf

1. Maintain and strengthen US Leadership in high-field accelerator magnet technology for future colliders;
2. Further develop and integrate magnet research teams across the partner laboratories and US Universities for maximum value and effectiveness to MDP;
3. Identify and nurture cross-cutting / synergistic activities with other programs (e.g. Fusion), to more rapidly advance progress towards our goals.

These three core vision elements provide focus and direction to the program, while guiding interaction with other DOE-SC offices, international partners, and industry, to further the mission of the MDP.
The overarching goals of the program remain unchanged after the program's first four years:

- ***Explore the performance limits of $Nb_3Sn$ accelerator magnets***, with a sharpened focus on minimizing the required operating margin and significantly reducing or eliminating training
- ***Develop and demonstrate an HTS accelerator magnet with a self-field of 5 T or greater***, compatible with operation in a hybrid HTS/LTS magnet for fields beyond 16 T
- ***Investigate fundamental aspects of magnet design and technology*** that can lead to substantial performance improvements and magnet cost reduction
- ***Pursue $Nb_3Sn$ and HTS conductor R&D*** with clear targets to increase performance, understand present performance limits, and reduce the cost of accelerator magnets

There has been significant progress towards these goals on multiple fronts since the inception of the US MDP. As examples, a record dipole field (14.5 T) has been produced by the FNAL cosine-theta magnet "MDPCT1" (see Figure 1); a Bi-2212 magnet has achieved 4.7 T in a common-coil configuration, without exhibiting training; and recent $Nb_3Sn$ developments have shown the potential to greatly enhance vortex pin density in $Nb_3Sn$ conductor architectures, for example by incorporating point pinning with $ZrO_2$ or by enhancing grain-boundary density through the addition of Hf, or some combination of these approaches. Using these techniques, new $Nb_3Sn$ prototype strands have demonstrated breakthrough properties for HEP applications [1] [2]. Finally, new diagnostics and data analysis techniques are providing unique insight into the magnet training performance characteristics.

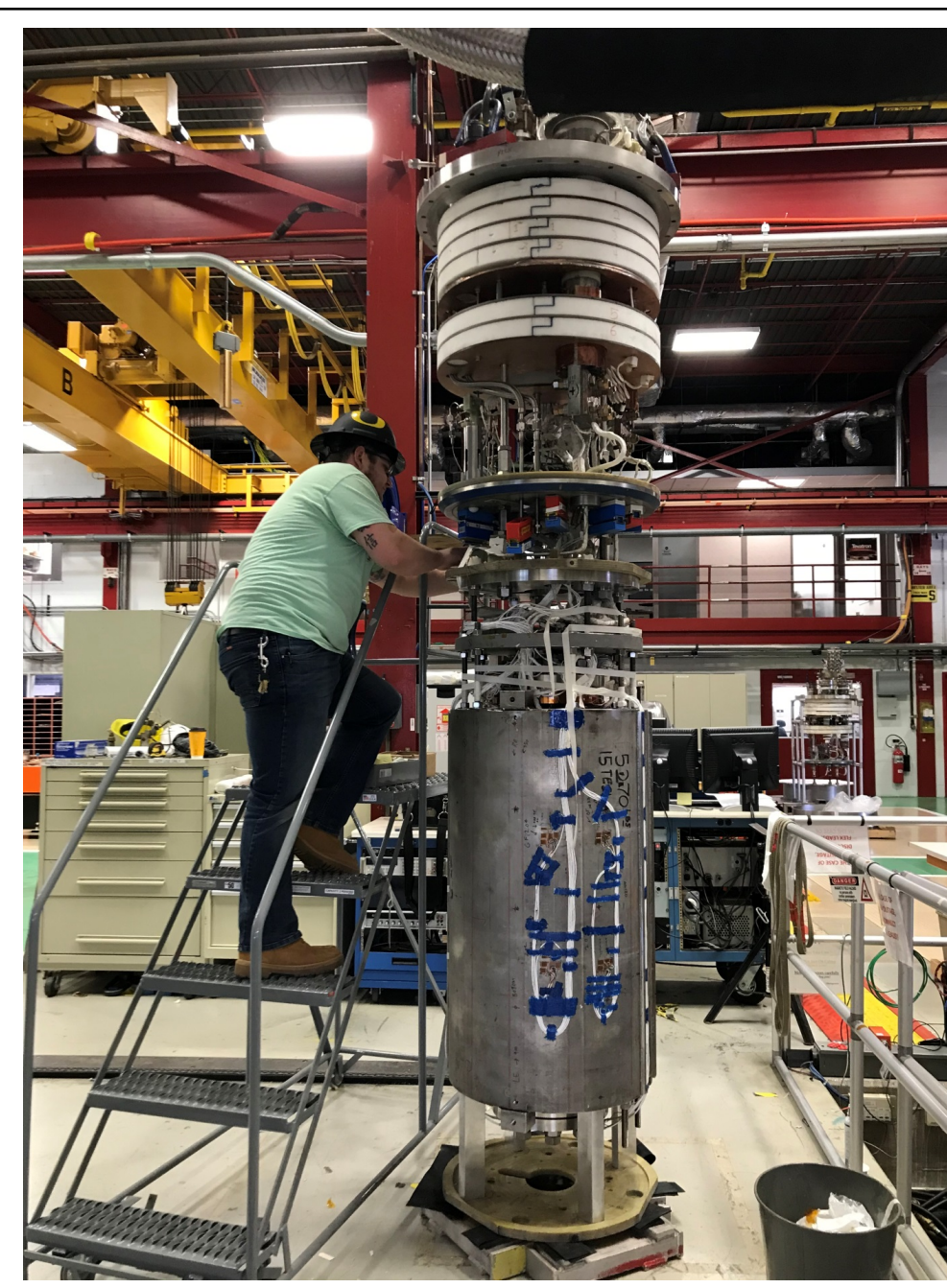
*Figure 1. The 60-mm aperture dipole demonstrator MDPCT1 being prepared to cold tests. The magnet reached the record field of 14.5T in June 2020.*

Guided by these developments, the updated roadmaps shift the MDP emphasis towards HTS magnet development, enhance investments in fundamental aspects of magnet technology, and expand development of hybrid HTS/LTS magnets. $Nb_3Sn$ superconductors and magnets nevertheless remain central to the program, playing a key role in our investigation into stress-management as a possible means to address forces at high field and in large-bore magnets, which must provide the background field for our hybrid HTS/LTS prototype magnet goals.

## Driving questions

High field accelerator magnet research is driven by questions related to ultimate magnet performance, magnet cost, industrialization feasibility, accelerator operation, and appropriate choice of superconductor. Overarching driving questions are shown in Table 1.

*Table 1. Driving questions for the US Magnet Development Program. These questions guide the program's near- and long-term goals, and serve to focus the program's research.*

| Q# | Driving questions |
|---|---|
| | **Ultimate Performance of Magnets** |
| 1 | *What is the nature of accelerator magnet training? Can we reduce or eliminate it?* |
| 2 | *How do we best define operating margin for $Nb_3Sn$ and HTS accelerator magnets, and to what degree can and should it be minimized?* |
| 3 | *Can we control the disturbance spectrum and engineer a magnet response to reduce operating margin and enhance reliable performance?* |
| 4 | *What are the mechanical limits and possible stress-management approaches for $Nb_3Sn$, HTS, and 20 T hybrid LTS/HTS magnets, and do they have defined mechanical limits?* |
| 5 | *Do hybrid designs benefit from the best features of LTS and HTS, or inherit the difficulties of both material technologies?* |
| | **Cost, Industrialization, and Operation** |
| 6 | *What is the optimal operating temperature for $Nb_3Sn$ and HTS magnets?* |
| 7 | *What are the possibilities and limitations associated with safely protecting $Nb_3Sn$ and HTS magnets?* |
| 8 | *Can we provide accelerator quality $Nb_3Sn$ magnets beyond 16 T? What are the operational field limits for $Nb_3Sn$ magnets?* |
| 9 | *What is the optimal operational field for $Nb_3Sn$ dipoles? For hybrid HTS/LTS dipoles?* |
| 10 | *Can we build practical and affordable accelerator magnets with HTS conductor(s)?* |
| 11 | *What drives the economics of high field accelerator magnets? Are there innovative approaches to magnet design that address the key cost drivers for $Nb_3Sn$ and HTS magnets and do they shift the cost optimum to higher fields?* |
| | **Superconductors for Accelerator Magnets** |
| 12 | *What are the near and long-term goals for $Nb_3Sn$ and HTS conductor development? What performance parameters in $Nb_3Sn$ and HTS conductors are most critical for high field accelerator magnets? Can we effectively define limiting factors (properties, cost, manufacture) of present HTS conductors and accelerate their development to industrial maturity?* |
| 13 | *Prototype HTS magnets made so far, whether made from Bi-2212 or from REBCO have not shown training even in dipole geometry where $Nb_3Sn$ is particularly sensitive. Is it possible to envisage NO TRAINING as a potentially vital, cost-saving attribute of HTS conductor use?* |

## Program structure

**Strategic directions for the update plan:**

- *Probing stress management structures*
- *Hybrid HTS/LTS designs*
- *Understanding and impacting the disturbance-spectrum*
- *Advancing both LTS and HTS conductors, optimized for HEP applications*

The program is structured to align with the overarching goals of the program:

I. A $Nb_3Sn$ magnet development effort is central to the program. Building on the successful test of the FNAL cosine-theta dipole magnet, and to mitigate increases in coil stresses at higher fields, the program will focus on developing stress-managed concepts. The Canted-cosine-theta (CCT) dipole effort, led by the LBNL team, will continue, with a near term focus on rapid-turn-around subscale prototypes that enable systematic development and proofing of the technology prior to scaling up to high field. A stress-managed cosine-theta (SMCT) dipole concept, recently developed and led by FNAL, that leverages traditional cosine-theta magnet technology will proceed in parallel. The SMCT approach complements the CCT approach, strengthening a major goal of the updated plan: to determine if stress-managed structures can fulfill their promise to break the traditional scaling of coil stress with field and truly enable high field magnet technology with strain sensitive $Nb_3Sn$.

II. HTS magnet development with parallel initiatives in REBCO and Bi-2212 magnet technologies remains central to MDP. Both efforts, originally based on the CCT and racetrack coil approaches, have made excellent progress over the last four years, and the original goal of achieving 5 T in a stand-alone configuration is close. All laboratories and University program collaborators are actively engaged in this HTS program. The ongoing CCT efforts will be reinforced by stress-managed cos-theta approaches for Bi-2212 and REBCO coils. A major theme for the next few years will be developing and demonstrating hybrid magnet operation for accelerator magnets.

III. Investigating the fundamental aspects of magnet technology – the "science" of magnets – has only grown in importance and emphasis within the program over the first four years. A plethora of new structural materials and fabrication techniques, diagnostic concepts, new data analysis methodologies, new modeling techniques, and exploratory ideas to improve training, are being pursued. Furthermore, a new element to the technology arena has been added, dubbed "Comparative analysis of magnet designs". It is focused on evaluating a broader spectrum of high field magnet designs to explore potential future directions for the program so as to provide early insight into the technical challenges of ~20 T hybrid magnet designs.

IV. Central to all high field accelerator magnet technologies is the conductor itself, and we intend to enhance our investments in conductor R&D, while in parallel providing sufficient commercial conductors for magnet construction to proceed. Conductor R&D has made significant strides since the initiation of the MDP, both through direct MDP investments and, importantly, through the alignment of strategy with other funding opportunities. Examples of progress include dramatic increases in current density in Bi-2212 wires via breakthroughs in powder manufacturing and in overpressure processing; steady reduction in REBCO conductor tape substrate thickness and significant increases in vortex pinning via Zr doping, leading to significant improvements in overall current density and in reduced bending radius of REBCO cables; the first demonstration of the potential for $ZrO_2$ particles and Hf doping to enhance vortex pinning in $Nb_3Sn$, and adding high-$C_p$ materials to composite $Nb_3Sn$ wires, breathing new life into the application of $Nb_3Sn$ for high field accelerator magnets.

## Technical area updated roadmaps

### Area I: $Nb_3Sn$ Magnets

There have been significant advances in accelerator dipole magnet development over the last couple of decades [3]. The 60-mm aperture, 15 T dipole demonstrator MDPCT1 (see Figure 1) developed by MDP reached 14.1 T at 4.5 K in its first test in June 2019 [4]. Subsequently the magnet was reloaded for higher field, and reached 14.5 T at 1.9 K in its second test in June 2020. Both fields are world records for accelerator magnets at these temperatures. Analysis indicates that further increase of the magnet's operational field, or increase of its aperture, will require coil stress management techniques to mitigate conductor degradation. The updated $Nb_3Sn$ magnet roadmap aims to develop and demonstrate the effectiveness of stress-managed (SM) approaches. The program focuses on a) increasing magnet operational field and aperture, b) reducing potential for magnet degradation, and c) reducing magnet training. The program furthermore considers means to minimize magnet cost. Stress management approaches are also critical for the HTS program elements due to the strain-sensitivity of HTS conductors and their ultimately intended higher field use, thus linking the aims of the $Nb_3Sn$ program to those of the HTS program.

Two complementary approaches are being pursued to investigate the potential for stress managed structures:

#### *Stress-Managed Cos-Theta (SMCT) dipoles*

The SMCT R&D goals are a) to develop and demonstrate a new approach to manage the radial and azimuthal stresses in brittle cos-theta coils, through the study and reduction of magnet training; b) to demonstrate a bore field up to 11 T at 1.9 K with 120-mm aperture [5] in two-layer $Nb_3Sn$ dipole magnets with stress-managed coils [6]; and c) to demonstrate up to 17 T at 1.9 K with a 60-mm aperture in a four-layer $Nb_3Sn$ dipole magnet with stress-managed outer coils [7]. Dipole cross-sections are shown in Figure 2. Major milestones for the SMCT program element are provided in Table 2.

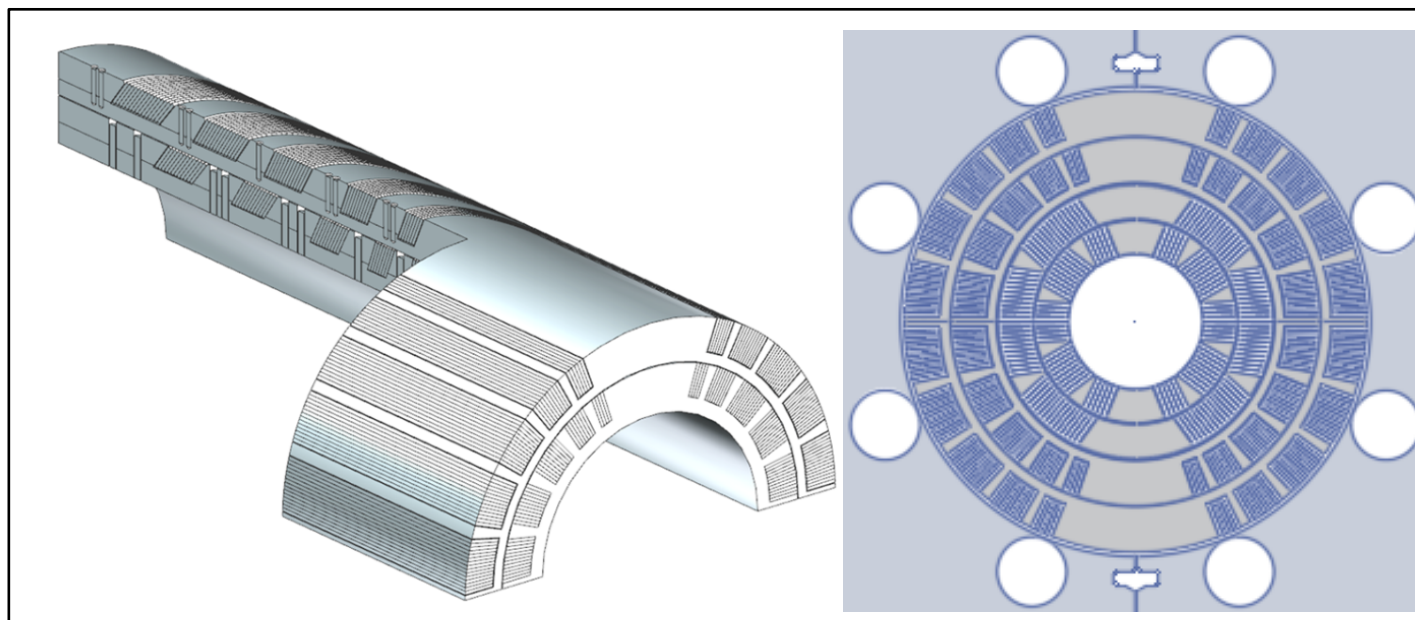

*Figure 2. 3D view of two-layer 120-mm aperture stress management coil (left) and four-layer 60-mm aperture $Nb_3Sn$ dipole coil cross-sections (right), that are part of the updated MDP roadmaps for $Nb_3Sn$ magnet development.*

### Canted-Cos-Theta (CCT) dipoles

The CCT R&D goals include a) further understanding of interfaces and design and fabrication methods for CCT/stress management using dedicated subscale tests; b) testing and improved understanding of novel instrumentation approaches through dedicated CCT subscale tests; c) pursuing improved modeling approaches of interfaces to further understand the performance of stress managed magnets; and d) demonstration of bore field up to 13 T in a 120-mm aperture in four-layer $Nb_3Sn$ CCT model magnet. An example CCT coil is shown in Figure 3. Major milestones for the CCT program element appear in Table 3.

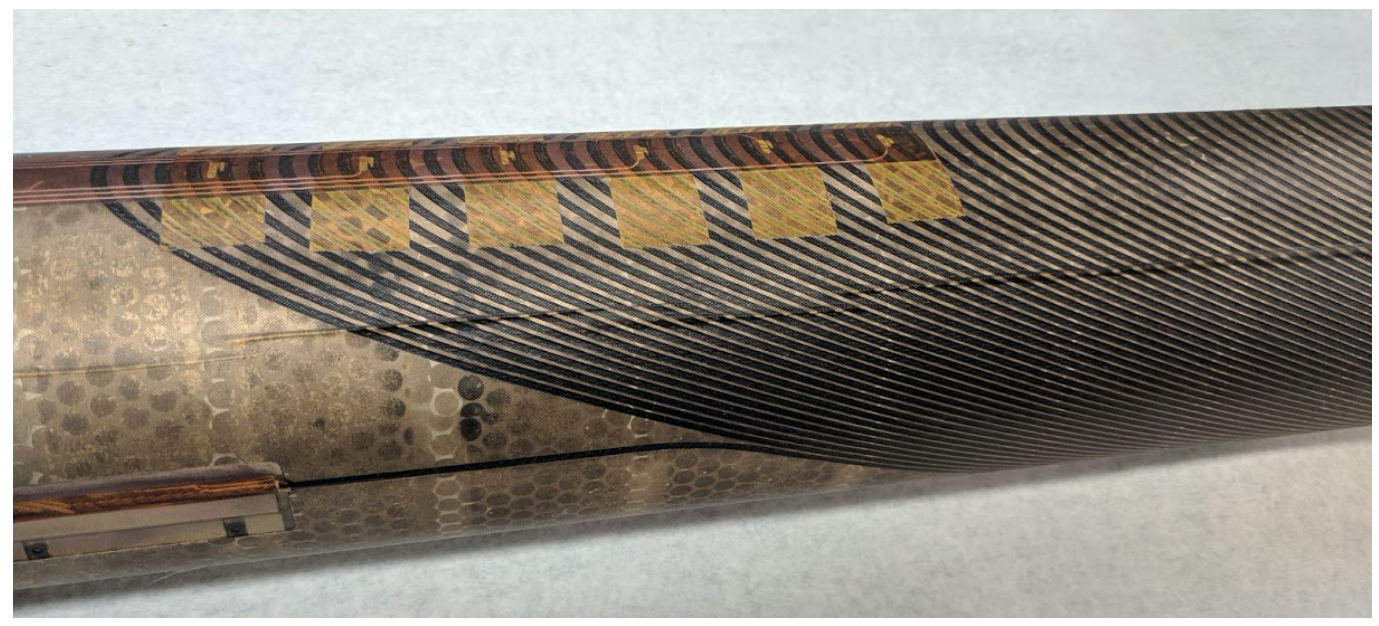

*Figure 3. A single layer of a $Nb_3Sn$ CCT magnet: each turn of the Rutherford cable is supported by a machined groove in the mandrel. A "trace", effectively a lithography patterned flexible sheet of Cu-coated Kapton, is seen on top, providing access to voltage taps on the turns of the magnet.*

### *Driving technical questions*

The two development platforms are designed to address major goals of the updated MDP plan:

1. Do "stress-managed" magnet designs deliver on the promise to mitigate the traditional increase in, and hence ultimate limitation of, stress with field strength and magnet bore?
2. Can "stress-managed" structures serve as the foundation for hybrid HTS/LTS magnets?

### General goals of the $Nb_3Sn$ area

The stress-managed (SM) magnet concepts described above address their corresponding US-MDP driving questions, formulated in Table 1, and provide the strong dipole fields and large bore needed for subsequent HTS insert coil tests. Integration of design, fabrication infrastructure, instrumentation, test facilities and test data analysis from the MDP partner labs will increase the efficiency and outcomes of the program. We note that a "utility structure", capable of accommodating these magnets as well as other hybrid magnets being developed by the program, has been designed over the last two years [5], but not yet fabricated. The program will determine the most effective mechanical structures for the SMCT and CCT magnets as the $Nb_3Sn$ magnet program evolves.

Through the pursuit of these two complementary approaches, the program will most efficiently and effectively probe the potential of stress-managed structures while maximizing the opportunity to develop a robust platform for hybrid HTS/LTS magnet development. The updated $Nb_3Sn$ program roadmap is shown at the top of the *Updated MDP Roadmap* in Figure 9.

## Area II: HTS magnets

The development of accelerator magnet technology using high-temperature superconductors (HTS) is a growing component of the Magnet Development Program. HTS materials continue to evolve, with major improvements in performance and growing industrial conductor manufacture. Nevertheless, conductors remain costly and there are significant technical challenges that must be overcome to arrive at a feasible technology for accelerator magnets. The updated MDP plan builds on our experience to date, and aims to integrate the HTS and $Nb_3Sn$ magnet developments through the use of hybrid HTS/LTS magnets that have the potential to achieve very high fields as efficiently and rapidly as possible.

### Area IIa: Bi-2212 magnets

Leveraging an academia, industry, and national lab partnership, US MDP in 2017 improved the $J_E$ of best Bi-2212 wires by ~60% to 1000 A/mm$^2$ at 4.2 K and 27 T [8]. Such improvement is a critical step towards developing this conductor for practical magnet applications in particle colliders. The wire performance improvement, together with new heat treatment and insulation materials, led to quadrupled quench currents in subscale racetrack coils made at LBNL using 17-strand Rutherford cables, compared to previous coils fabricated during the Very High Field Superconducting Magnet collaboration in 2009-2011 [9]. Their training-free quench behavior is also intriguing. The racetrack coils assembled into a common coil dipole magnet that achieves 4.7 T in a 6 mm gap. Promising performance was also demonstrated in prototype coils based on the CCT design [10].

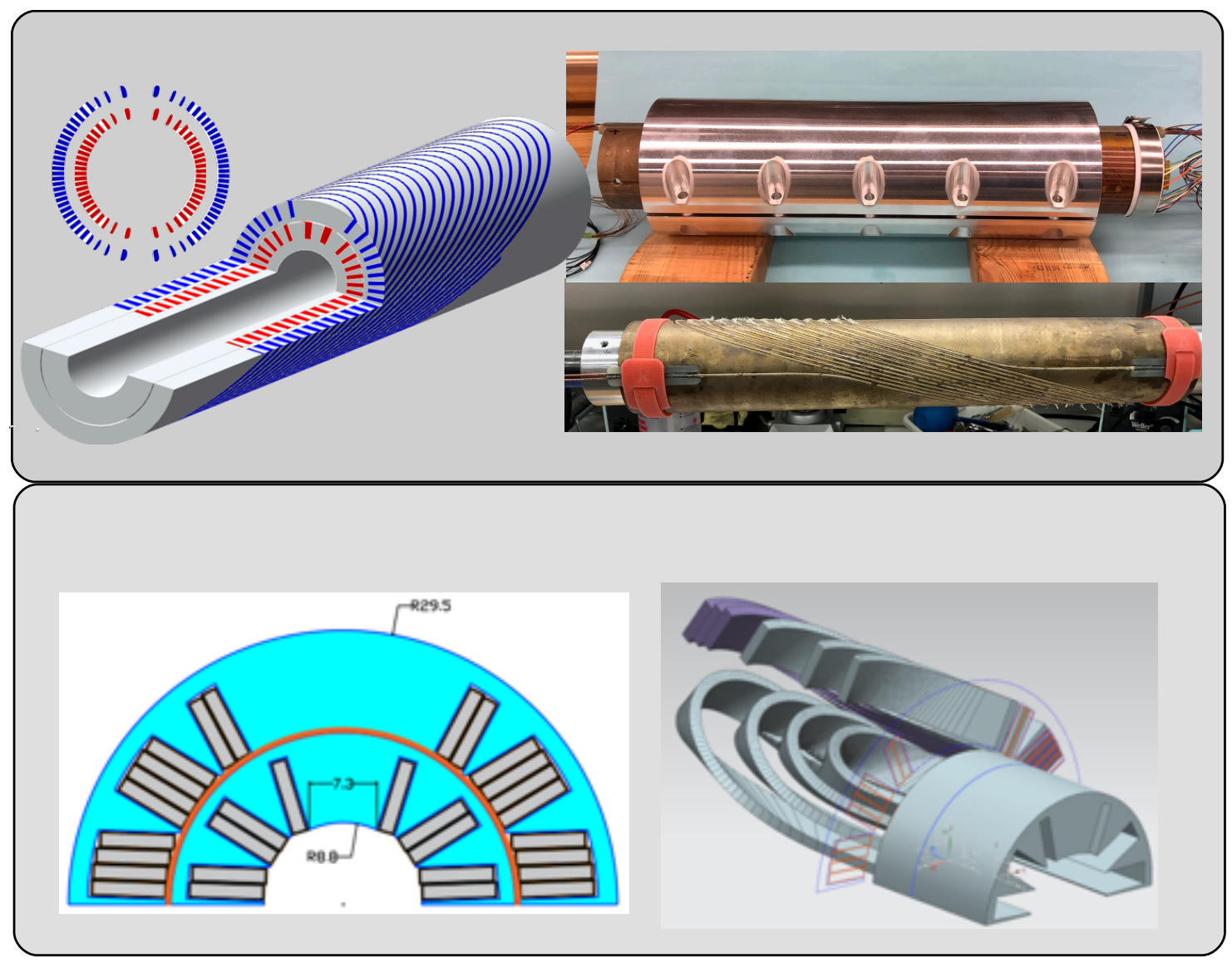

*Figure 4. Bi-2212 insert magnet concepts that will be explored in MDP. Top: CCT concept(left) and first implementation, showing reacted coil and the final magnet ready for testing. Bottom: Stress-managed cosine-theta concept, showing cross section and 3D CAD model.*

The natural next steps are to capitalize on the improved conductor performance in accelerator magnets and exploring applications beyond particle colliders. The initial prototype CCT B-2212 coils indicate that a Bi-2212 CCT has the potential to become a high-field accelerator magnet technology. The US MDP intends to construct Bi-2212 CCT magnets with 50 mm bore, measure and model their field quality, with gradual improved field generation (2.4 T, 3.5 T, and 5 T). The US MDP will also explore the SMCT concept for Bi-2212 [11]. The SMCT work will be focused on the 17-mm aperture 2-layer Bi-2212 SMCT dipole coils with the design self-field of 5.5 T. This target field will be approached gradually by using the old and new generations of Bi-2212 wires. The Bi-2212 magnet concepts are shown in Figure 4.

Both CCT and SMCT coils will be tested in the background fields of $Nb_3Sn$ magnets being developed within the MDP Area I (see above). These R&D vehicles will help us to understand key questions related to Bi-2212 magnet design and technology.

### *Driving technical questions*

1. Can high-field CCT and SMCT Bi-2212 magnets with good field quality be reliably manufactured?
2. What is the maximum field for training-free Bi-2212 CCT and SMCT coils?
3. What are the field generation and stress management limits of CCT and SMCT Bi-2212 magnet technologies?
4. How do we protect $Nb_3Sn$/HTS hybrid magnets from quenches?

The MDP Bi-2212 coil program will leverage the ongoing facility development at the NHMFL to expand their present overpressure heat treatment facility (Deltech 140 mm bore by 400 mm long in which all recent MDP coils have been reacted) with a new furnace (RENEGADE) that is capable of processing Bi-2212 coils of 1 m length and up to 250 mm in diameter. There is strong synergy in between MDP and the NHMFL in this area (see section *The National High Magnetic Field Laboratory* below). Furthermore, the US MDP will continue the conductor R&D collaboration at the NHMFL with wire and powder industries to further improve Bi-2212 conductors.

### Area IIb: ReBCO magnets

The high-temperature superconducting $REBa_2Cu_3O_{7-x}$ (REBCO, RE = rare earth elements) conductors have an unmatched capability to carry high current density over a wide range of temperatures and magnetic fields, giving them a special potential for broad applications. Of particular interest to MDP are high-field magnets generating a dipole field of 20 T and above for future circular particle accelerators. A highly synergistic application is compact fusion reactors aimed at early commercial electricity generation. About ten manufacturers in the U.S., Europe, and Asia are competing and producing commercial REBCO tapes but real understanding of their various properties is still lacking.

Compared to accelerator magnet technology for other advanced conductors such as Bi-2212 and $Nb_3Sn$, the REBCO accelerator magnets require significant technology development and demonstration on multiple fronts: multi-tape cables, magnet design and fabrication, and reliable quench detection and protection [12]. REBCO tapes are supplied in the superconducting state, which is advantageous compared to $Nb_3Sn$ and Bi-2212 that require high-temperature and overpressure heat treatment.

The REBCO magnet development will focus primarily on utilizing CORC® or similar cables in two stress-managed configurations, the CCT concept and a novel "Conductor on Molded Barrel" (COMB) concept (see Figure 5) [13]. To help the MDP effectively exploit the significant potential of REBCO conductors, we list the following driving technical questions that need to be addressed for the REBCO program area.

*Driving technical questions*

1. *What performances and architectures should the conductor have? How can we make magnets using these conductors?*

High-field accelerator magnets gravitate towards multi-tape cables in order to reduce inductance and facilitate magnet protection and operation. The strain-sensitive cables have implications for safe magnet design and fabrication that will minimize any strain-induced degradation. Magnet design and fabrication will help guide the conductor development: architecture, transport performance, bending radius, inter-tape contact and etc. Impregnation and joint fabrication will also be addressed.

2. *What is the maximum dipole field a REBCO magnet can generate? What is the long-term conductor and magnet performance under cycling Lorentz loads?*

The stress/strain limit of REBCO conductors will largely determine the ultimate field a REBCO magnet can generate. The performance of REBCO conductors under cyclic Lorentz loads is an important factor to determine the long-term magnet performance.

3. *What is the magnet performance? How does it quench? How can one protect the magnet during a quench? What is the field quality of REBCO accelerator magnets?*

Magnet performance provides the best feedback for conductor and magnet technology development. Can we exploit a stable flux flow transition to devise a safe detection and protection strategy? Side effects of the large REBCO current density and the wide-tape, single-filament geometry of a coated conductor are strong magnetization and screening currents, whose impact on machine performance need to be understood.

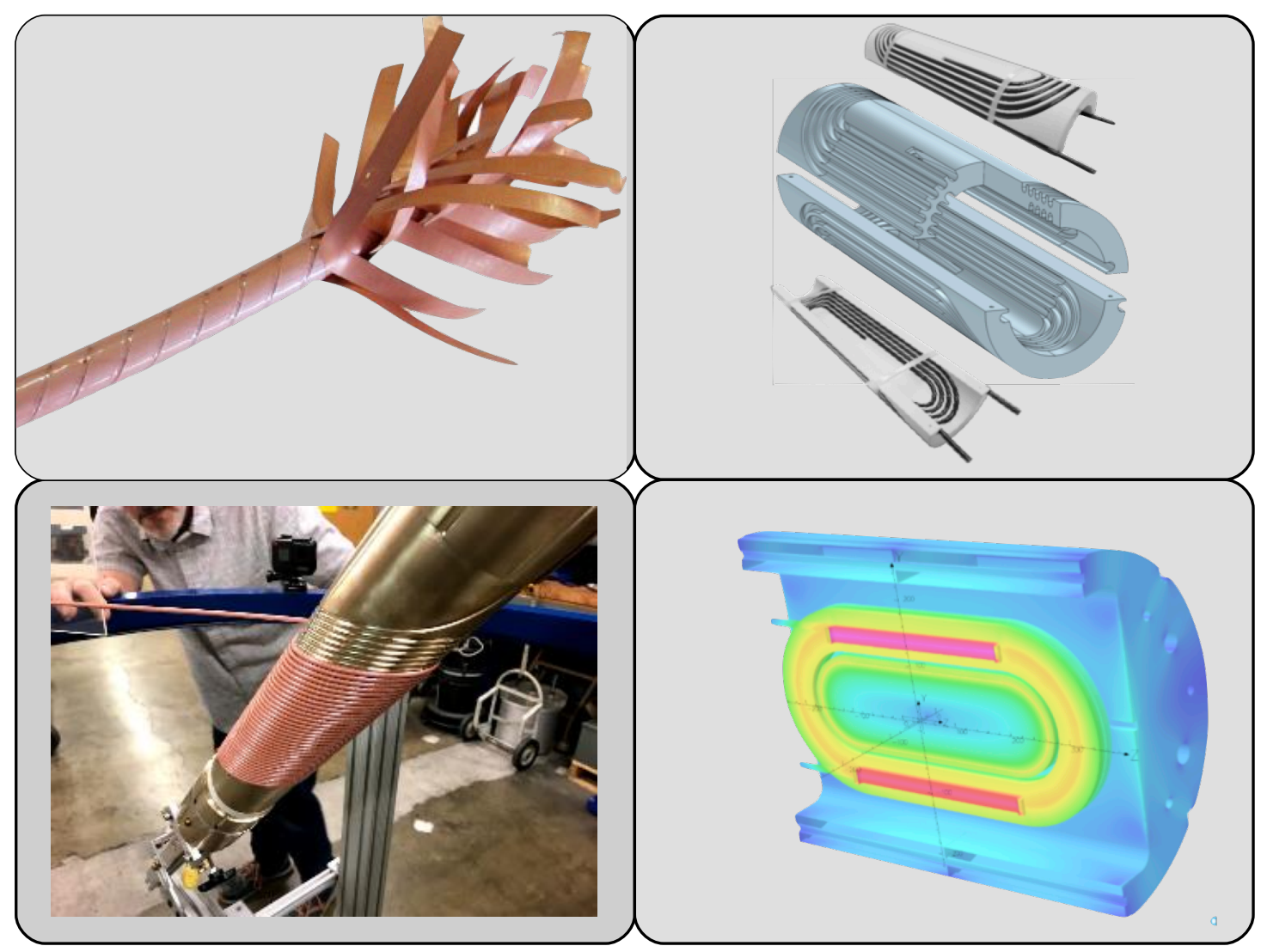

*Figure 5. Elements of the REBCO program. Top left: CORC® wire from Advanced Conductor Technologies. Top right: Novel "COMB" concept for stress management [13]. Bottom right: first test of REBCO racetrack tape-wound coils tested in BNL's 10T common coil magnet. Bottom left: winding of "C2", a CORC-based CCT dipole.*

4. *How can we characterize the transport performance for long conductors?*

Future magnets will require cabled, multi-strand conductors with unit piece lengths of 100 – 1000 m. REBCO tapes come with pre-existing manufacturing defects and may develop further damage during the cabling process. How do we locate these defects in long magnet conductors before magnet fabrication?

The MDP will explore the potential of REBCO for accelerator magnets through the use of stress-managed structures that help mitigate the potential for conductor damage from transverse forces and resulting stresses. The CCT approach, along with a novel "Conductor on Molded

Barrel" (COMB) concept, will serve as the initial platforms for REBCO coils compatible with hybrid testing in $Nb_3Sn$ outserts developed in Area I.

## Area III: Technology

### *Overview*

In support of the magnet development areas of the program, a plethora of science and technology developments are pursued by the US MDP. To provide structure and focus to these broad areas, we have organized the Technology Area into five subtopics (a-e), and each subtopic team has developed plans and milestones that address critical needs of the magnet sections of the program. The corresponding roadmap elements form the last section of Figure 9.

### Area IIIa: 20 T hybrid magnet design and comparative analysis

Superconducting magnets for particle accelerators with operation fields beyond the 15-16 field level, currently considered a limit for $Nb_3Sn$ superconducting technology, will have to rely on coils with HTS conductors. HTS conductors open the possibility of increasing the field to 20 T (or higher), but, taking into account the higher cost of HTS with respect to LTS conductors, a hybrid solution, where both LTS and HTS are utilized, has to be considered and investigated. A hybrid magnet constitutes a significant challenge in particular from the point of view of mechanical integration of different coils, testing and protection. The goal of this task is to perform a design of a 20 T hybrid magnet and address the following challenges.

#### *Driving technical questions*

1. What are the mechanical limits and possible stress management approaches for 20 T hybrid LTS/HTS magnets?
2. Do hybrid designs benefit from the best features of LTS and HTS, or inherit the difficulties of both materials?
3. Is there a design option (CT, block, common-coil) more suitable for hybrid magnets?
4. What drives the cost of 20 T accelerator magnets? How can it be minimized?
5. Can we build practical and affordable accelerator magnets with HTS conductor(s)?

#### *Goals*

The task has two main goals. The first is to carry out a comparative analysis of different design options for a 20 T hybrid magnet which utilizes both LTS ($Nb_3Sn$) and HTS superconducting materials. In particular, the following lay-outs will be investigated:

I. Cos-theta design and its stress management option
    a. "Traditional" cos-theta (CT) design
    b. Canted cos-theta (CCT) design
    c. Stress-management cos-theta (SMCT) design
II. Block-type coil design (block coil with flared ends)
III. Common-coil design (block coil with racetrack coils)

The study will start with the definition of key design criteria (like operational field, load-line margin, field quality, stress levels, etc.) and will be focused on the specific challenges of integration of LTS and HTS coils, protection, powering, and test facility set-ups for hybrid magnets. 2D analytical and finite element models will be used to perform the conceptual design; in addition, 3D considerations will be accounted for in the comparison between the different design options. Other general but key objectives will be to review the strengths and weaknesses of the different designs and provide inputs to other areas (in particular Area I and II on LTS and HTS magnet developments) to better define their roadmaps and avoid inconsistencies.
The second goal will be to review and follow-up current work on hybrid magnets (with field levels ranging from 9 to 14 T), collect and analyze the data gathered within this activity, and use the data as an input for the design of a 20 T hybrid magnet.
Key milestones spanning a 3-year timeframe are presented in Table 6. Throughout the 3 years, results from low-field hybrid tests will be collected, reviewed and used as input for the analysis of the 20 T hybrid.

### Area IIIb: Advanced modeling

Advanced modeling tools are currently utilized across the full range of US-MDP research activities, enabling the design of improved conductors, magnets, and diagnostics. The updated program plan places a strong emphasis on stress-managed designs and hybrid HTS/LTS magnet testing. This focus leads to a unique set of technical challenges that, as part of an integrated effort with other US-MDP areas, can be addressed with new modeling tools and techniques.

#### *Driving technical questions*

The goals outlined above are designed to address the following driving questions related to magnet technology:

1. How does interface behavior in stress-managed designs impact training?
2. How do we best protect HTS/LTS hybrid magnets?
3. Can we better understand and then improve HTS/LTS conductors and cables?
4. How can we best interpret test results and diagnostics data?

Goals of the advanced modeling efforts include:

- model interfacial debonding and other advanced mechanical interface conditions to interpret training results;
- improve magnets through an “interface by design” effort which guides technology development in epoxies and impregnation techniques;
- model complicated, coupled quench behavior of HTS/LTS hybrid magnets [14];
- evaluate the effectiveness of existing quench protection schemes and develop new techniques targeting the specific needs of hybrid magnets;
- develop computationally efficient methods for simulating HTS that leverage DOE high performance computing resources;
- optimize HTS cables for accelerator applications through the understanding of quench, current sharing, and mechanical limitations;

- refine our understanding of the strain limits of $Nb_3Sn$ and HTS magnets.

### *Updated roadmaps and major milestones*

The updated roadmap for the advanced modeling area is shown in Figure 9. Major milestones are provided in Table 7.

### Area IIIc: Magnet Materials

Magnet insulation and engineering materials are a vital component in superconducting magnet fabrication. Insulation materials and processes are used to define the operating limits in magnet design, from voltage standoff limits to mechanical degradation limits. As magnets reach their design points, the microscopic behavior of these materials dominates the magnet performance, manifesting as training behavior, and perhaps sets ultimate limits on achievable fields in practical applications.

Development of improved magnet insulation materials and a whole magnet approach to selection and material design is anticipated as a way to substantially expand the achievable parameter-space of high-performance magnet designs. Figure 6 shows an example measurement system that probes critical mechanical properties of magnet components.

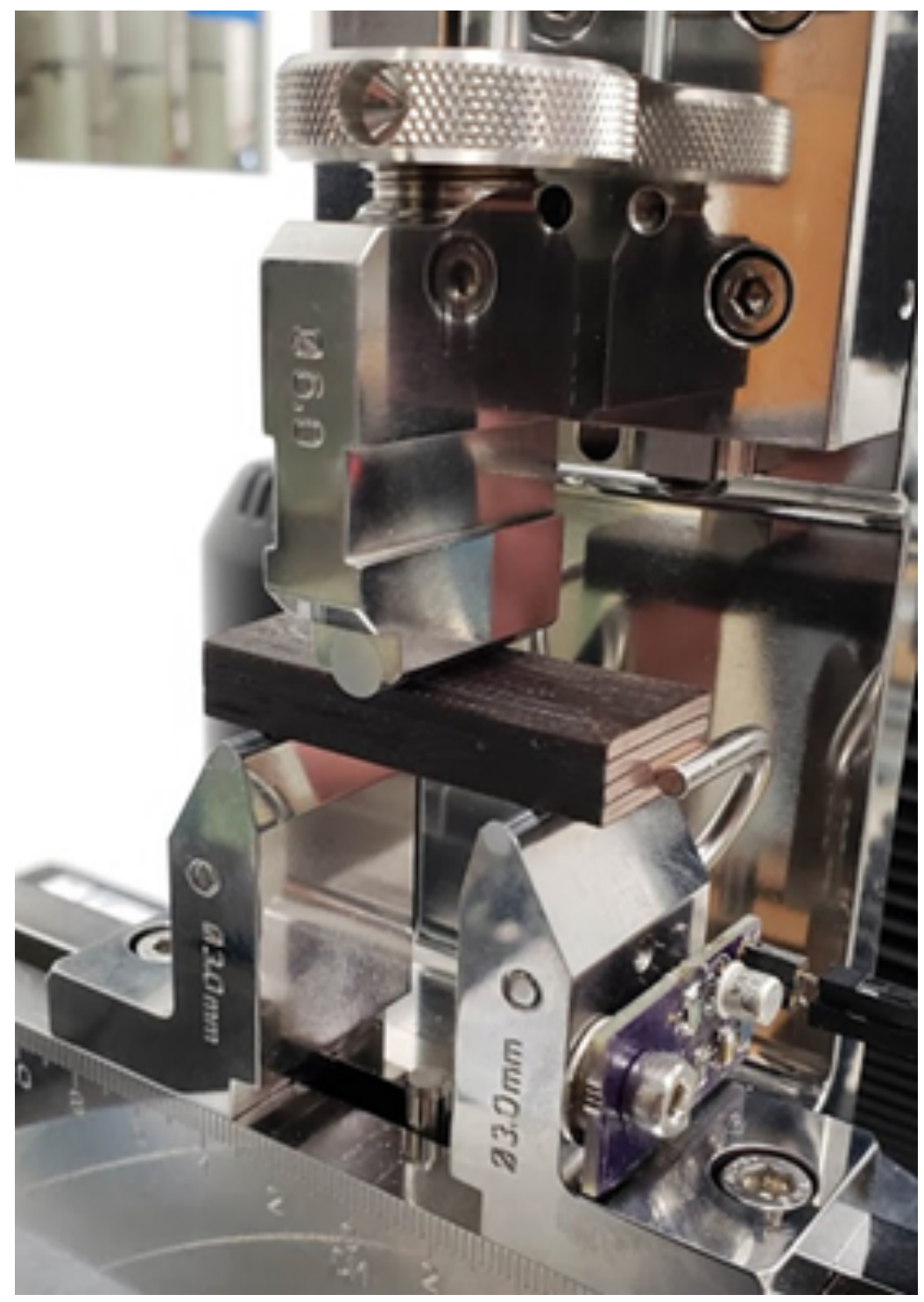

*Figure 6. A custom "ten-stack" mechanical measurement system at FNAL to probe compression and shear modulus, delamination strength, and other properties.*

The interface between different classes of coils in a hybrid magnet is not the only interface of concern. Different magnet impregnation materials exhibit different behavior on different substrates. Bonding and interface behavior internal to a coil is of major concern, as small cracking may release enough energy to quench a ($Nb_3Sn$) magnet, leading to extended training or limited achievable current. These considerations motivate a focus on tailoring interfaces through chemical or mechanical means to design behavior needed to enable practical, efficient magnet designs.

Materials investigations are intended to determine the interaction and behavior of components of a magnet and may allow behavior to be tuned through system formulations, preparation methods, or topology improvements, or other modifiable parameters. Coupled with Advanced Modeling, these investigations can enable magnet optimization.

### *Driving technical questions*

1. How do magnet materials directly affect magnet training and performance?
2. How can magnet materials be designed to facilitate improved magnet performance?

3. Can improved processes lead to more cost effective and robust magnet designs?

The general approach is to identify specific parameters and/or processes that may impact performance, develop techniques/protocols, and demonstrate with subscale composite characterization; then proceed with a down-selection and a subscale magnet fabrication.
*Initial milestones intended to enable new technologies and exploration of new magnet designs are provided in*

Table 8. The experience will then be integrated into the magnet elements of the program, and will guide further investigations moving forward, with additional materials inputs expected from SBIR and other programs.

## Area IIId: Novel Diagnostics

Magnet diagnostics are an essential and rapidly advancing component of the program. They provide a unique "observation window" into mechanical and electromagnetic processes associated with magnet operation, quenching and training. They help identify intrinsic performance-limiting factors in magnets and provide essential feedback to magnet design, simulations and material research activities. A broad spectrum of novel diagnostic approaches is presently being explored by the program. Acoustic sensors leveraging advances in piezo-sensors and compact cryogenic amplifiers allow direct measurements of elastic energy release during magnet ramping. A precise timing of the acoustic wave arriving at multiple sensors enables 3D localization of mechanical disturbances and quenches in complex magnet systems with an accuracy of few centimeters. Acoustic spectrograms are used to "fingerprint" magnet mechanical events using machine learning techniques. Diffuse-field ultrasonics allows for a non-invasive detection and localization of hot spots in HTS coils and conductors. Magnetic quench antennas and Hall sensor arrays enable mapping of current redistribution, conductor instabilities and quench development in LTS and HTS magnets. New techniques such as fiber-optics and capacitive sensing aiming at a local real-time monitoring of magnet strain and temperature are being actively explored. Finally, novel analog and digital electronics for liquid helium temperature operation are being designed and tested to enable a new generation of diagnostic instrumentation. Our plan for the next three years focuses on developing and implementing novel sensor hardware, electronics and data analysis for real-time, non-invasive monitoring of LTS and HTS magnets.
We will use several diagnostic techniques in synergy to access the physics of quench-triggering disturbances and study mechanisms of mechanical memory and training in $Nb_3Sn$ magnets. We will also develop unique non-invasive tools to image current-sharing patterns in superconducting cables, localize hot spots and achieve robust and reliable quench detection in HTS coils and hybrid LTS/HTS magnets.

### *Driving technical questions*

1. How do we resolve and properly identify mechanical and electromagnetic disturbances in magnets and understand the physics of the training process?

2. How do we non-invasively localize weak points and interfaces where mechanical disturbances leading to premature quenching take place? Can we manipulate those interfaces *in situ* to improve magnet performance?
3. How do we achieve a robust, reliable and minimally invasive quench detection and localization for HTS magnets? Can we practically realize a new paradigm of HTS magnet operation where quenching can be avoided altogether through an early detection?
4. How do we resolve current sharing patterns and stress-driven defect accumulation in HTS coils and cables to ensure their robust long-term operational stability and ultimate quench resilience?
5. Can we advance magnetic field measurements to a next level by implementing arrays of miniature magnetic sensors combined with computationally-advanced field reconstruction algorithms?
6. Can we drastically simplify diagnostics instrumentation while making it more efficient and reliable using cryogenic electronics, in particular FPGAs and quantum sensors?

A list of milestones for this subsection of the Technology area are provided in Table 9.

### Area IIIe: Training Reduction

One of the common features of $Nb_3Sn$ accelerator magnets is their long training characteristic needed to achieve fields close to 80-85% of the fields permitted by the conductor properties.

#### *Driving technical questions*

To explore the performance limits of $Nb_3Sn$ conductors in accelerator magnets focusing on minimizing the required operating margin and significantly reducing or eliminating training (MDP Goal 1), we pose the following questions:

1. What is the nature of accelerator magnet training, especially in $Nb_3Sn$ magnets?
2. Can we find mechanisms to reduce or eliminate the training?

#### *Technical approaches*

A number of potential techniques aiming to affect the training curve before or during magnet powering are under investigation:

- High-Cp conductor and insulation development that can lead to conductors and coils with optimized characteristics enabling stable operation against perturbations [15];
- Artificially increasing the coil current during a quench by discharging a large capacitor at quench detection. The "overcurrent" transient will generate forces beyond the nominal quench current, possibly significantly increasing the rate at which the magnet trains and hence reducing the total number of training quenches;
- Inducing ultrasonic vibrations into the conductor and coil parts during ramp up. This may allow for gradual energy discharges, avoiding accumulation of energy (notably due to friction) in areas around the coil/conductor and potentially improving the rate at which magnets train;

- Performing fast-turn-around cable/stack training R&D in a controllable configuration, thereby providing more quantitative insight into training mechanisms and means to influence them.

These approaches complement other technology areas that are focused on understanding the disturbance spectrum and on mitigating their sources. Milestones for this subsection are provided in Table 10.

## Area IV: Conductor Procurement and Research & Development (CPRD)

### Area goals

The CPRD mission is threefold:

1) Maintain a viable inventory of conductor for foreseen magnet R&D
2) Invest in developmental magnet conductors
3) Define needs and opportunities for University, SBIR, and LDRD programs

To carry out this mission, CPRD operates an advisory committee, whose role is to review proposals from conductor manufacturers and needs from MDP magnet teams, and then recommend funding actions by MDP for purchase of production conductor, investment in developmental conductor, investment in other research activities, and direction of conductor in inventory toward specific uses. In FY 2020, CPRD aims to identify procurements and investments of approximately \$0.5 to \$1.0 million value annually. CPRD aims to maintain this level of investment for FY2021 and beyond. Importantly, MDP has adopted a management strategy to set aside conductor funds, based on a recommendation from external review.

“Production” refers to conductor delivered according to a purchase specification with guaranteed performance. For $Nb_3Sn$, production quantities are 40 kg billets and procurements can be well in excess of 100 kg, where 5 kg of conductor at 0.85 mm diameter yields about 1 km length. “Developmental” refers to conductor delivered according to a best-effort agreement, with expectations for properties and length but no guarantees. Typical billet mass for developmental conductor is 10 kg. Bi-2212 strand and innovations of $Nb_3Sn$ are presently being produced at this scale, which would yield sufficient wire to permit manufacture of short cables and small coils. Other research activities generally have funding from sources other than CPRD, and can include more basic research and feasibility studies at a few kilogram scale. When appropriate, especially exciting results can be kick-started by CPRD.

The CPRD Advisory Committee membership is provided in Table 16. Review of CPRD Advisory Committee membership coincides with the review of the MDP program. CPRD acts in a nexus of other activity that includes University-driven research, laboratory-directed R&D (LDRD), early career research awards (ECRA), small-business innovative research (SBIR), and broader programs at CERN and US national laboratories. CPRD has a leading role in supporting industrial development at the major manufacturers, which are not eligible for SBIR awards, and CPRD also has a supporting role to facilitate the interaction between emerging ideas from University, LDRD, and ECRA activities and major manufacturers. Figure 7 presents a diagram of this nexus.

The CPRD team extends the Conductor Development Program (CDP) originally developed in the mid-1990s [16]. CDP was an early supporter of Bi-2212 development and also contributed largely to the development of advanced $Nb_3Sn$ magnet conductors while working in conjunction with LARP, where contributions by CDP played a central role in defining the conductor architecture, alloying choices, heat treatment strategies, property trade-offs, and limitations for use for accelerator magnets now being manufactured for the HL-LHC upgrade [17]. Following that tradition, CPRD will work closely with MDP thrust areas to understand the needs of magnets under development and the technology opportunities and trade-offs that advanced conductors could provide.

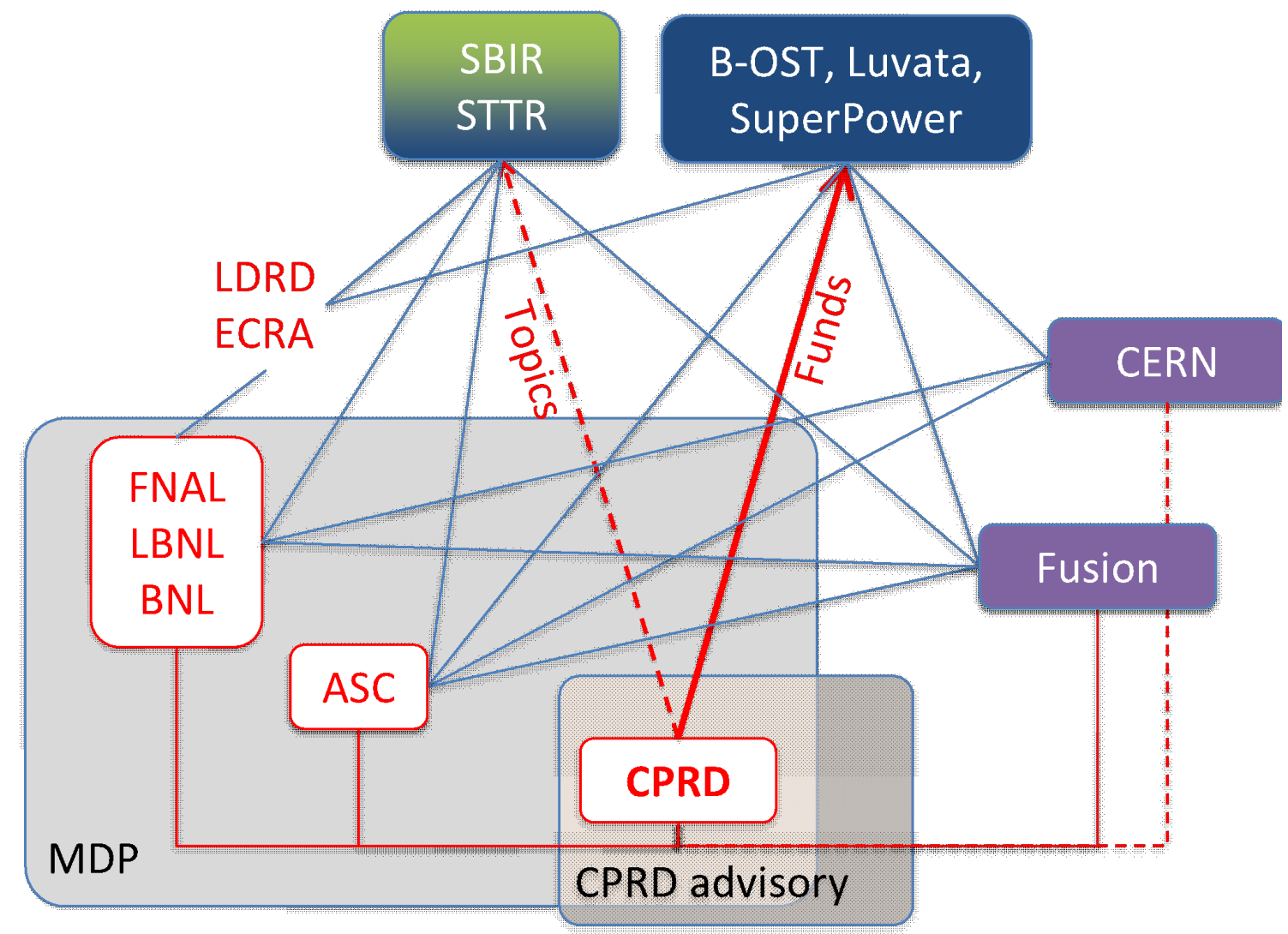

*Figure 7 Nexus diagram of different activities relevant to MDP and CPRD objectives. Here the box labeled ASC identifies the Applied Superconductivity Center, but it is representative also of programs at other universities with support from DOE that are not formally affiliated with MDP.*

## Roadmaps and major milestones

### *$Nb_3Sn$*

A variant of the $Nb_3Sn$ conductor presently in mass production (~10 tons purchased by US labs) for the HL-LHC forms the baseline conductor for MDP. Magnet designs call for conductor with 0.7 mm up to 1.3 mm diameter, which can be supplied as 127, 169, or 217 element re-stacks with most components, e.g. 108, 150, or 198, being $Nb_3Sn$ sub-elements. Manufacturing standards can be maintained while increasing the tin component slightly in comparison to the HL-LHC conductor. New reaction strategies, along with the extra tin, allow conductors to exceed 1400 $A/mm^2$ non-copper current density at 16 T, 4.2 K, which is about a 15% increase over the capability of the HL-LHC conductor.

Opportunities for significant improvement in current density are presented by the advent of Zr- and Hf- doped alloys. These alloys are effective in two regards:

1) When oxygen is supplied, e.g. by including tin oxide in the conductor architecture, pinning centers are formed in the $Nb_3Sn$ layer, e.g. $ZrO_2$ particles, which augment the pinning by grain boundaries.
2) When high-melting point Hf is alloyed, the recrystallization temperature is increased above the reaction temperature used to form $Nb_3Sn$. The retention of fine grains facilitates a reduction of the $Nb_3Sn$ grain size and stronger vortex pinning and higher $J_c$.

The approaches above could add 30% or more to the critical current density at the field range of interest to MDP.

Opportunity to add conductor stability is also afforded by addition of high heat-capacity material, such as rare-earth oxides and rare-earth borides, to the conductor architecture. Enhanced heat capacity has been shown to increase the minimum quench energy threshold, which is relevant to magnet training.

Roadmap: Over the next 5 years, CPRD aims to assist in the scaling from present lab and University research program levels (few kg) to industrial development level (~10 kg) and possibly to industrial production level (~40 kg). Proposals will be solicited for efforts to scale individual approaches, e.g. Hf-alloyed material, and combined approaches, e.g. combination of pinning additive with high heat-capacity additive, along this path.

### Driving technical questions – $Nb_3Sn$:

1. What, if anything, is required to tailor the present production conductor for a magnet operation target in the 16 T range?
2. Can the opportunities for Zr- and Hf- doped variants with added pinning centers and refined A15 grain size be realized in production-level conductors at reasonable cost?
3. Do high heat-capacity additives have a valuable role, and if so can they be incorporated into production-level conductors at reasonable cost?

### *Bi-2212:*

CPRD and NHMFL procurements, as well as conductor purchases by other entities not connected with MDP, have established an initial experience base that could facilitate statistical tracking of future development. Billets are typically 10 kg, which yield about 1 km of conductor at 1.3 mm diameter and 2 km of conductor at 0.8 mm diameter. The former parameters are preferred for solenoids at NHMFL, while the latter is associated more with cables made at LBNL. Success has been obtained with a variety of architectures, where typically 85 x 18 and 55 x 18 restacks have been used. Recent procurements have been delivered in single pieces as well as 5 to 7 pieces.

Powder raw material has stabilized with Engi-Mat as a primary supplier. Up until 2020, production lots have been 2 kg, i.e. a 10 kg billet above combines powder from 5 batches, with new scale-up to 10 kg powder batch production. Two metrics of powder quality are emerging:

1) The peak critical current obtained for a short-sample 50 bar over-pressure heat treatment (OPHT), where University teams vary the reaction maximum temperature $T_{max}$ and the time in melt $t_{melt}$ to locate an optimum: for 0.8 mm diameter CPRD strand received between 2017 and 2020, this parameter has ranged from 600 to 700 A, with one champion value of 900 A, at 4 K and 5 T (note: the 4 K, 15 T critical current scales as 66% of the 4 K, 5 T value, and for comparison a HL-LHC $Nb_3Sn$ strand attains 400 A at 4 K, 15 T).
2) The short-sample 50 bar OPHT averaged across a $T_{max}$ window of 10 °C. For the strand above, this parameter has ranged from 450 to 550 A at 4 K, 5 T.

Also, powders are screened for particulate size and evidence of chunks or agglomerates. At the present time, wire breakage appears to have a connection with chunks and agglomerates in the powder.

Roadmap: Two priorities are envisioned by CPRD. First, the present best-effort development conductor is ready to mature toward a production-like conductor with quality control and specifications. Second, implementation of conductor in coils and magnets will inevitably expose new risks and unknowns that will instigate conductor improvement. Accordingly, the roadmap milestones incorporate a program of steady investment in conductor billets while maintaining flexibility to adapt investments according to emerging needs.

### Driving technical questions – Bi-2212:

1. What are the challenges confronting an eventual production process, now that two US manufacturers produce high-quality powder and several magnet activities are using strand?
2. What must be done (on the conductor side) to get *coil* reactions to achieve the most out of the conductor?
3. What actions can be taken to reduce cost?

### *REBCO Coated Conductor:*

The LBNL-led activity to implement conductor on round core (CORC) cables made from REBCO coated conductor established a number of benchmarks from which CPRD can define development paths:

- Production of 400 m of REBCO on 25 µm Hastelloy, with slit widths down to 1.5 mm
- Critical current of >350 A at 6 T, 4.2 K, which demonstrates successful incorporation of pinning additives comparable to the production level for thicker material
- CORC strands wound on 2 mm core at 90 m length, using 30 coated conductors above and achieving > 5 kA at 6 T, 4.2 K .

Roadmap: Experience with the first CORC coils identified many potential areas for CPRD investment with an overall goal of developing a REBCO-based conductor compatible with HEP accelerator magnet needs. The updated roadmap is focused on delivering the requisite conductor and developing and demonstrating the magnet technology in both standalone and hybrid HTS/LTS configurations.

Driving technical questions – REBCO:

1. Can challenges of REBCO coated conductor manufacturing be overcome? Challenges include: slitting cracks, variations in REBCO thickness and texture, variations in vortex pinning center density, short (< 200 m) pieces, variations in contact resistance between conductors, tolerance of drop-outs and flaws that arise during use.
2. Can production characteristics for thicker coated conductors, including advances to overcome challenges above, be achieved in the specially produced thin (20–25 µm) and narrow (1–2 mm) conductors for HEP?

## Roadmaps for the US MDP

The 2020 Updated MDP Roadmaps have been developed in two forms. First, a high-level ten-year roadmap is shown in Figure 8. The plan is designed to align with the US community planning process ("Snowmass") and the anticipated Physics Project Planning and Prioritization Process ("P5") that will follow. We have additionally identified high field solenoids as a *possible* area of future focus within the MDP, contingent on community support, strategic future facility need, and DOE-OHEP support.

Second, a detailed 2020 Updated MDP Roadmap has been developed, focused on the next 3-5 years (see Figure 9). The detailed roadmaps were generated by the integrated MDP team through a series of dedicated meetings, reviewed, and presented for feedback, at a dedicated international workshop held in Gaithersburg, Md. in December 2019, and finalized at the MDP Collaboration meeting held in Berkeley in February 2020. The detailed roadmaps are consistent with the MDP Area descriptions outlined previously.

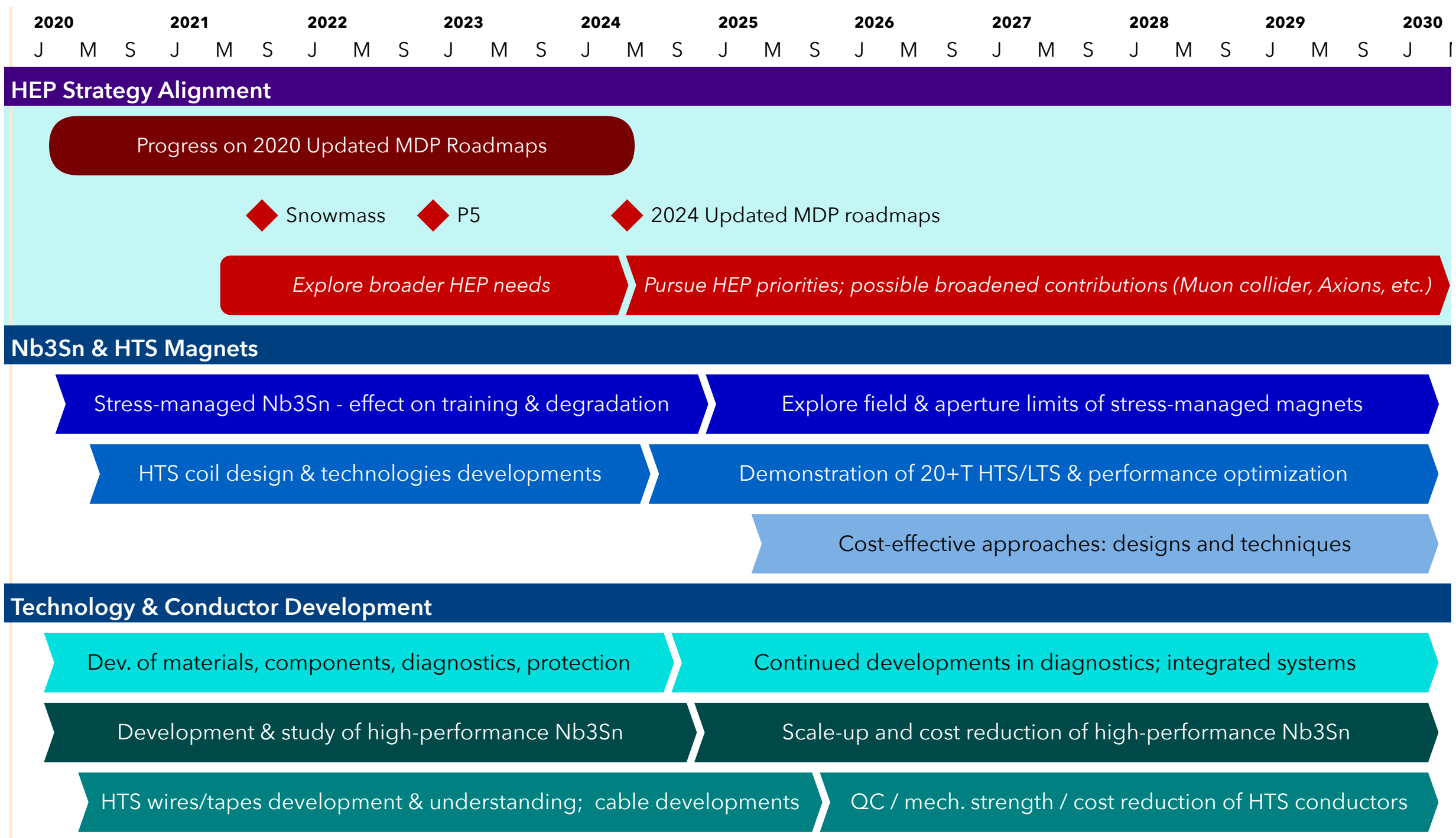

*Figure 8. A high-level ten-year roadmap for the US Magnet Development Program, designed to align with the US Physics community planning process.*

J

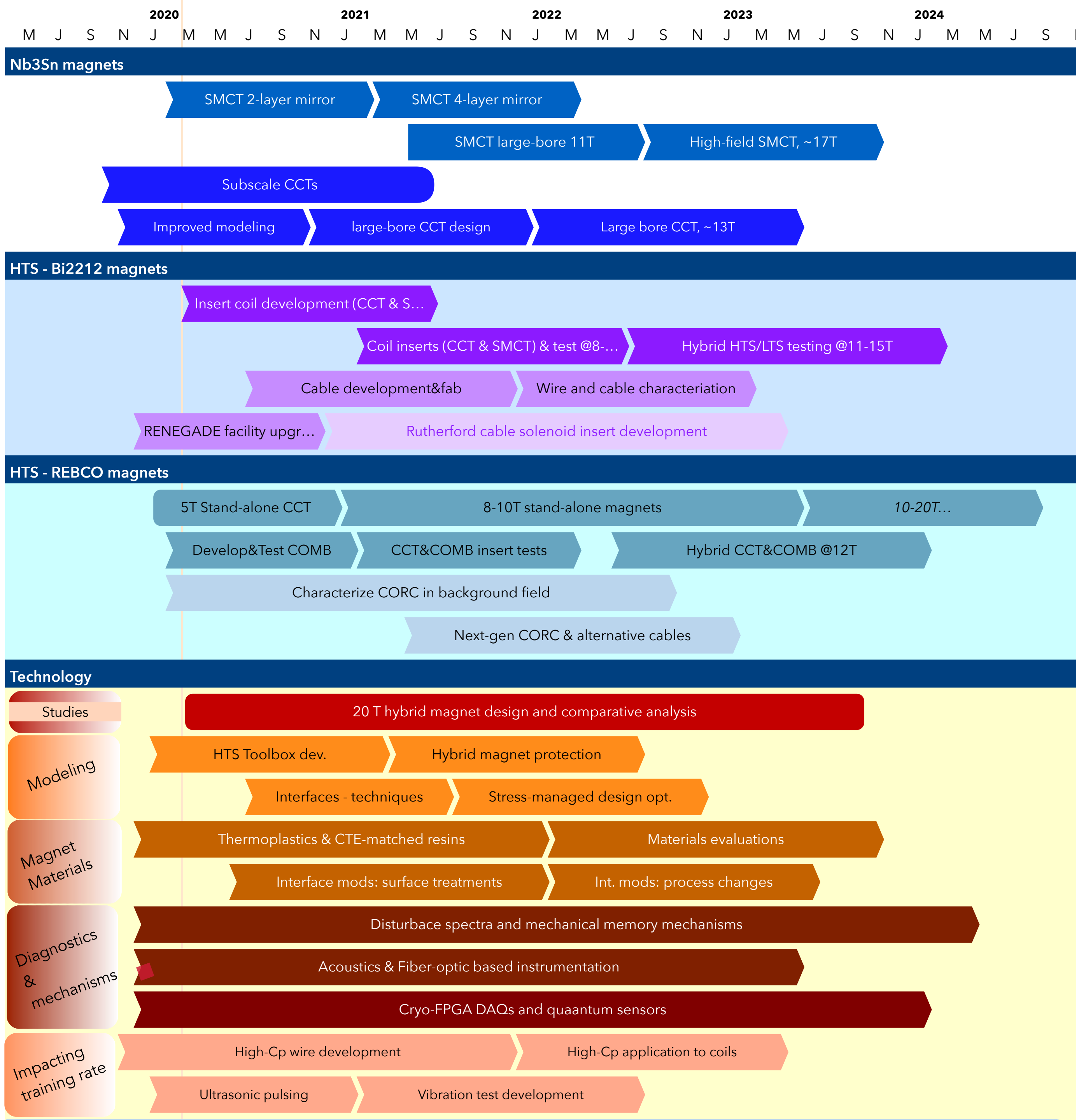

*Figure 9. The updated roadmaps for the major elements of the program, including $Nb_3Sn$ Magnets, HTS (Bi-2212 and REBCO) Magnets, and the various Technology areas. The $N_{b3}Sn$ magnet designs will focus on stress-managed structures, motivated by the need to intercept forces in magnets at high field and with large bores compatible with hybrid (HTS/LTS) configurations.*
*Work within the MDP on HTS magnets to-date has focused on stand-alone magnets; the focus will now shift towards hybrid magnets, with first tests expected in 2021.*
*Nomenclature: "SMCT"=Stress-managed cosine-theta; "CCT"=Canted cosine-theta; CORC®=Conductor-on-round-core, trademark of ACT Inc.; "CTE"=coefficient of thermal expansion; "FPGA"=Field-programmable gate array; "DAQ"=Data acquisition system; "High-Cp"=material doped to enhance heat capacity.*

## Prioritization process

The roadmaps outlined above provide a comprehensive suite of research elements designed to deliver major and lasting advances in accelerator magnet technology for High Energy Physics applications. To address the realities of funding and resources that can be allocated for the program, a process is in place to guide priorities so as to most effectively deliver on the program goals. Central to the updated roadmaps are the following driving elements:

- Development of "stress-managed" $Nb_3Sn$ large bore high-field magnet(s) that both evaluate stress management as a concept, and provide background field for hybrid testing. Note that the program investigates two complementary paths to this effect, the CCT and the SMCT. Depending on budget and on performance characteristics identified during development, either both lines can yield magnets for hybrid testing, or we can down-select to one that is then shared across the program.
- HTS insert development based on CCT, COMB and SMCT design concepts. The HTS magnets are designed to be compatible with testing in a $Nb_3Sn$ outsert. The inserts are tested in stand-alone mode to understand and characterize HTS coil behavior. As above, depending on budget and on insert coil performance, either both lines can be used for hybrid testing, or we can down-select to one that is then shared across the program.
- Hybrid testing of HTS/LTS magnets, i.e. installing the HTS inserts in the aperture of large-bore stress-managed $Nb_3Sn$ magnet(s). The goal is to provide early feedback on the viability of hybrid magnets, critical to guide future program emphasis of HTS magnet technologies.
- Development of advanced conductors ($Nb_3Sn$ and HTS) and demonstrating their relevance to magnet technology. This includes demonstrating performance compatible with magnet application (stability, cabling, scalability in length, etc).

The Technology section of the program has many elements, and some prioritization is needed to guide the level at which those elements can be supported; that process will be dictated in large part by their anticipated impact on addressing/supporting the above driving elements. This is best done "organically", with feedback and guidance from the magnet elements of the program. Another element that will be folded into the prioritization of technology elements is the recognition of the importance, in particular for junior staff, of seeding innovative concepts to provide opportunities for breakthroughs and for career development. Innovation must be central to the MDP – it is the hallmark of HEP's longstanding vision and support for general magnet R&D. Naturally other funding mechanisms to support innovation will also be explored to enhance the program's effectiveness where possible.

In the area of $Nb_3Sn$ magnets, the program intends to prioritize magnet development and the understanding of magnet performance over magnet "demonstrators". As an example, the 17T $Nb_3Sn$ magnet, nominally planned as the combination of the large-bore 2-layer stress-managed $\cos(\boldsymbol{\theta})$ surrounding the existing 15T magnet inner coils, will be pursued if at that point in the program it is deemed an effective means of answering key questions from Table 1. Certainly testing inserts made of advanced $Nb_3Sn$ wires in the same 2-layer SMCT will have a high priority, should cables made from such wires become available.

Finally, the program intends to prioritize hybrid magnets over dedicated stand-alone HTS magnets in the near term (we note that HTS coils designed for hybrid magnet testing are of course tested in stand-alone configuration as part of their qualification). This is motivated by two considerations: first, HTS magnets developed by the MDP to-date have exhibited very intriguing and promising performance characteristics, such as training-free behavior. Hybrid magnets offer the program the most rapid and cost-effective means of evaluating HTS magnet performance at higher field. Second, although hybrid magnets offer a form of conductor "grading" that in principle should yield the most cost-effective and compact magnet designs, the jury is out on the viability of the concept; it is important for the long-range program strategy to understand early if the concept is viable. We note that all HTS inserts produced for hybrid testing will be tested in stand-alone mode first, as part of their qualification and characterization.

## Synergistic programs and activities

### The National High Magnetic Field Laboratory

The NHMFL is the premier high field magnet lab in the world and its mission is to provide high magnetic field access to more than 1000 users per year. Although the work horse user magnets in the DC program in Tallahassee are principally resistive magnets using 20-34 MW per magnet and supplying fields up to 45 T DC, there are superconducting magnets, presently up to 20 T, for users who wish to spend extended periods at high fields where a resistive power bill expense is insupportable. For this reason, the NHMFL always took a strong interest in high field superconducting magnets.

Two National Research Council reports, COHMAG[2] (2004) and MagSci[3] (2013) have described the scientific and technology rationales for many new types of ultra-high field (UHF) superconducting magnets: regional 32 T superconducting (SC), national 40 T SC, 28–37 T SC high-resolution NMR, 25–40 T SC for x-rays and neutrons, 60 T hybrid DC, 20 T human MRI, as well as magnets for fusion, particle-accelerators, radiotherapy, axion and other particle detectors. The NHMFL, supported by the NSF, keeps these grand challenges in its mission statement. It has had an NSF-supported Science Driver aiming at the background R&D needed to fulfill these goals since 2007, first emphasizing conductor characterization, but recently placing greater emphasis on UHF test coils made from still-not-fully-developed High Temperature Superconductors (HTS). The world-record 32 T User Magnet[4] and 45.5 T insert test magnet [18] both used single-tape REBCO coated conductor (CC). Both magnets put their conductors under extreme stress and high energy density quenches. Central thrusts of ASC-NHMFL-supported work have been to:

---

[2] https://www.nap.edu/download/10923
[3] https://www.nap.edu/download/18355
[4] https://nationalmaglab.org/magnet-development/magnet-science-technology/magnet-projects/32-tesla-scm [18]

1. Develop a thorough understanding of HTS conductors, their $J_c(\boldsymbol{\theta},H,T)$ characteristics and their defect populations so as to be able to assess their uniformity and predictability, and their likely behavior under the high stress and quench conditions of the UHF magnets that would fulfill COHMAG and MagSci goals. The NHMFL has developed some special characterization capabilities (e.g. YateStar, an in-line 77 K, 0.6 T perpendicular and parallel to the tape plane transport measurement tool with integrated Hall probe array) in addition to its very high field magnet measurement capabilities.
2. Work on Bi-2212 round wire has largely been supported by DOE-OHEP through the open University annual competition process but the scope and interactivity of our work has been hugely enhanced by industrial interactions with companies like Nexans, Engi-Mat and MetaMateria and Bruker-Oxford Superconducting Technology (B-OST) and the magnet group at LBNL. The Over Pressure Heat Treatment (OPHT) process was developed out of a detailed study of current-limiting mechanism in Bi-2212 supported by DOE-OHEP. The role of NHMFL support was to stimulate this program in the 2007-2009 time period and then to shift support to manufacture of small coils including a 3 T insert inside 31 T that first demonstrated the benefits of OPHT for achieving high $J_E$ in coil forms. A large OPHT furnace was then purchased with NHMFL funds and with this high field NMR solenoids have been reacted as well as all recent LBNL racetrack magnets. A larger furnace suitable for 1 m dipoles has been jointly supported by both DOE-OHEP and the NHMFL. All of the essential elements needed for a wind-and-react (W&R) technology for Bi-2212 have been developed within the NHMFL-LBNL collaboration. Area IIa of the MDP depends vitally on these OPHT capabilities and ongoing technical collaborations. A significant FNAL collaboration on Bi-2212 is now also planned. The NHMFL-supported effort is now focused on high field NMR starting with ~1.2 GHz solid state NMR system made with Bi-2212 coils inside a Bi-2223 coil inside an existing large bore (212 mm, 12 T) LTS system.
3. The main thrust of the NHMFL HTS user magnet program has been REBCO coated conductor since this can be wound directly in solenoidal form as single tapes, both with insulated (I) tape geometry and in the so-called No Insulation (NI) magnet configuration. One of our R&D successes was driving an NI REBCO insert coil to a new DC field record of 45.5 T [18], however, not without significant damage to the coil. YateStar was vital to the *post mortem* of the damage and it has also been used to find both manufacturing and winding damage with CORC cables in collaboration with ACT and LBNL personnel. Understanding damage thresholds and mitigating measures for REBCO-based cables is a major element of MDP focus in Area IIb.
4. We have continued to work on understanding $Nb_3Sn$ conductors during the whole of the last 20 years and most recently have advocated for the addition of Hf or Zr to Nb4at.%Ta alloy to enable a much finer grain alloy that resists recrystallization during $Nb_3Sn$ reaction, thus raising $J_c$ to FCC levels. Although internal oxidation can also do this, it may be that the alloy grain refinement route is more compatible with standard internal-tin architectures like RRP from which HiLumi magnets are being made. This work is supported both by DOE-OHEP and CERN and industrial alloys have been procured from US industry (ATI and HC Starck).
5. Fe-base superconductors (FBS) are also being studied under DOE-OHEP support. There is potential for FBS conductors to have cheaper raw material and fabrication costs (at least in PIT form) than any other Nb-base or HTS conductor. With some FBS compounds having Tc of 35-40 K, $H_{c2}$ of order 90 T and an anisotropy close to 1, they are in many ways the "dream" future conductor for all future accelerators. A key question is whether the current flow obstacles seen to-date at higher field are really at grain boundaries as most believe and, if so, whether the obstacle is intrinsic as in the HTS cuprates or extrinsic, perhaps due to impurities at grain boundaries for which there is some evidence. Some recent work on FBS bulks and wires has started also at Fermilab under an LDRD and some joint experiments are in progress.

The overall NHMFL goals are to push HEP- and NHMFL-relevant conductors to their maximum, to make HTS test solenoids exceeding the next frontier of 50 T, and to work with MDP magnet groups to help all push back and finally understand the limits of the LTS and HTS technologies that underpin accelerator magnets. It may be worth noting that $Nb_3Sn$ was discovered to be the first high field superconductor in 1960, 60 years ago. Five or seven years ago when FCC issued its grand challenge of a $Nb_3Sn$ conductor with $J_c$(16T, 4.2K) of 1500 $Amm^{-2}$, almost no one thought there was any more development to do on $Nb_3Sn$. But the $J_c$ target has now been achieved by two groups within MDP by two different routes. This shows the benefits of the highly interactive collaborative-competitive culture nurtured by USMDP.

## Fusion Energy Sciences

There has been a longstanding interconnection between High Energy Physics and Fusion applications in the area of superconducting magnet technology. As an example, HEP has driven advances in $Nb_3Sn$ conductor performance, to the benefit of both fields; fusion has driven $Nb_3Sn$ conductor industrialization, through the massive procurements for ITER. This interconnection is greatly enhanced with the development of HTS superconductors and their potential to dramatically impact both HEP and FES applications. For fusion Tokamaks, the strong scaling of energy density with field provides enormous impetus to develop high field magnet technology that can enable compact, and hence ultimately cost effective, fusion reactors with high power production (see for example [19]). Once again, HEP research is driving conductor performance, and Fusion - and most notably private ventures in the field - is driving conductor scale-up and (presumably) cost reduction.

The strong synergy between HEP and FES in developing HTS conductor and magnet technology is manifesting itself on multiple fronts. Most prominently, the two programs are investing jointly in a common HTS Cable Test Facility, to be hosted at FNAL. The facility, nominally slated to come online in ~2024/2025, will provide up to 15 T field over ~750 mm length in a 100 mm x 150 mm test-well with a variable temperature insert. The design, fabrication, and testing of the 15 T dipole for the facility is being led by LBNL. In parallel, FNAL is preparing the site and associated infrastructure to host the magnet. Experiments will include $J_c(B,T,\boldsymbol{\varepsilon})$ measurements of high current cables critical for both applications. The new pit and vertical cryostat designed to host the Test Facility Magnet will also serve to test future hybrid HTS/LTS magnets being developed by MDP.

In addition, FES is investing in conductor and magnet research that builds on MDP developments, but tailors developments to fusion applications; the research is highly coordinated with MDP and in many cases leverages, and enhances, expertise and facilities from HEP. The FES program is actively engaged in MDP planning and collaboration meetings.

# Collaborations

## US Universities and Laboratories

Strong collaborative ties with US universities and laboratories provide a large pool of "virtual" resources, the best example being the NSF and DOE funded FSU/NHMFL program that is an MDP partner. Another prominent example is the Ohio State University program, which has longstanding support from HEP and contributes to the MDP on multiple fronts, primarily in conductor development and characterization. We note that the University programs are particularly productive in developing future scientists and engineers for conductor and magnet development; many graduates from the programs are now active members within MDP, within the superconducting industry, or have careers in international laboratories with close ties to the program.

The MDP Roadmap provides implicit guidance to other programs both domestically and abroad that helps to ensure focused and efficient progress on program goals. Our mutually beneficial collaborative activities extend beyond OHEP to other DOE programs such as OFES and ONP.

## International Universities and laboratories

All areas of the MDP benefit from close communication and collaboration with magnet and conductor R&D partners outside the US. Well-established communication channels ensure that we maximize progress through programs that are competitive but complementary.

Ties with international laboratories and universities have been, and we expect will continue to be, a critical element of our program. The US MDP management strives to foster a diverse, inclusive, and innovative culture that motivates innovation and open communication. Transparency in our purpose and research approach has proven to be effective in supporting the development of strong collaborations with international partners.

Furthermore, we note that a significant fraction of MDP staff were educated, wholly or partially, at international institutions. Exchanges of ideas, concepts, and results with international collaborators is central to our approach to research.

## Participation and Coordination with Global Strategic Planning Activities

The MDP plays a significant role in the development of one of the key enabling technologies for future particle accelerators and works closely with both national and international High Energy Physics planning activities. The organization and membership are actively engaged in the current Snowmass process[5] that will lead to the next Particle Physics Project Prioritization Panel (P5). The program also participated in the recent European Strategy for Particle Physics[6] (ESPP). The recently released ESPP update stated, *"the particle physics community should ramp up its R&D effort focused on advanced accelerator technologies, in particular that for high-field superconducting magnets, including high-temperature superconductors."* As the world leader in accelerator magnet R&D, the MDP will give the US an opportunity for a significant leadership role in a future collider regardless of geographic location.

[5] https://snowmass21.org/

[6] https://europeanstrategy.cern/home

## Industry

The MDP has strong connections with industrial partners, particularly in the areas of conductor and technology development. Our technical experts are heavily engaged in collaboration with industry partners in developing $Nb_3Sn$, Bi-2212 and REBCO conductors. A similar level of work is in the utilization of these conductors in a variety of magnet configurations through the development of new techniques for quench detection, magnet protection and cryogenic electronics that will aid the commercialization of magnets and other technology based on high performance low temperature and high temperature superconductors.

The MDP has very strong ties to small business, especially with complementary SBIRs and utilizing technologies developed as part of these SBIRs. There are many examples of SBIRs that directly or indirectly relate to MDP research. The common coil test stand at BNL that is currently being used to carry out magnetization measurements of YBCO coils and cable measurements was initially developed as part of an SBIR project with Particle Beam Lasers (Figure 10). In addition, some of the first hybrid HTS/LTS magnet tests were carried out as part of another SBIR also with Particle Beam Lasers. Furthermore, Advanced Conductor Technologies has also won an SBIR that is utilizing the common coil testbed to test a coil made of CORC conductor. BNL has partnered closely with ACT to develop a coil that will then be tested in the common coil test-bed. Another example of SBIR partnerships that develop novel technologies for the accelerator magnet community is the overpass /underpass coil. A proof of concept model was developed as part of an SBIR with e2P. Recently, another SBIR has been awarded to Particle Beam Lasers to work with BNL to develop a react-and-wind $Nb_3Sn$ overpass/underpass coil (see Figure 11) that will be designed to be tested in the common coil magnet as a corrector coil concept.

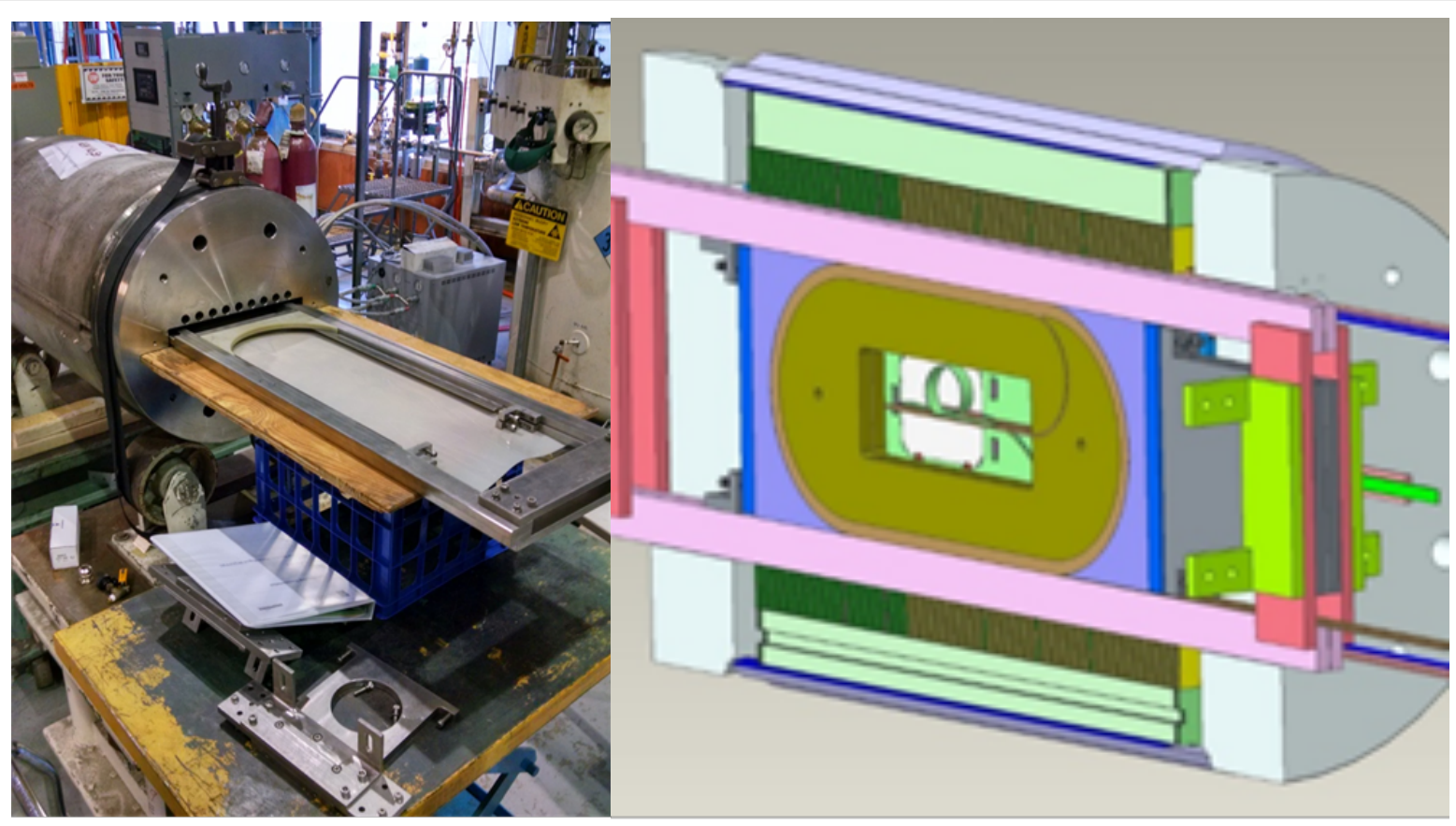

*Figure 10. Left: Common coil test stand at BNL. This test stand was developed as part of an SBIR with particle Beam Lasers. Right: An HTS insert coil being placed into the common coil test setup. This hybrid test was also funded as part of an SBIR with Particle Beam Lasers.*

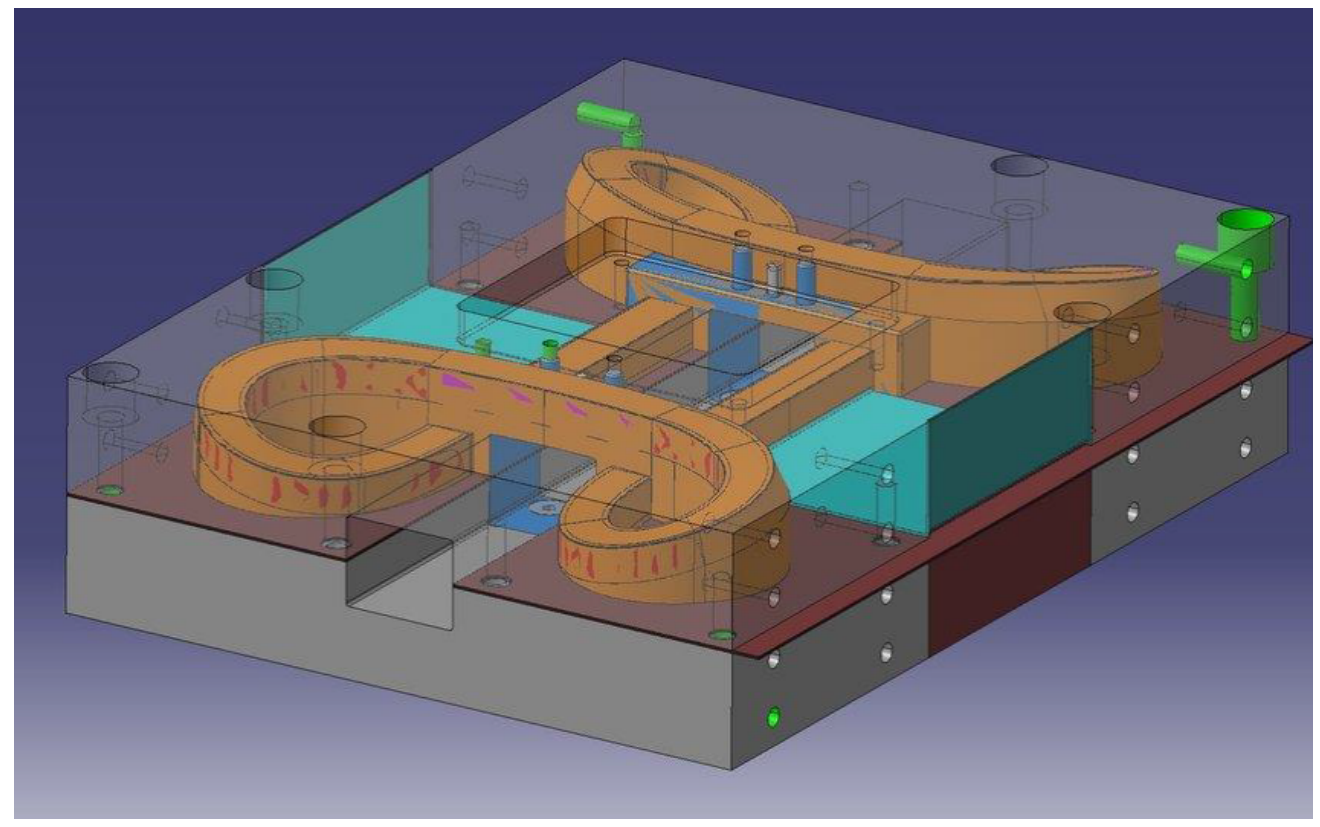

*Figure 11. Solid model of an overpass/underpass coil [20]. The concept, originally developed at BNL, is designed to accommodate strain-sensitive materials such as HTS or reacted $Nb_3Sn$, and is currently being explored by industry in an SBIR program.*

Multiple examples of SBIR's directly contributing to MDP goals exist in the materials realm. HyperTech has teamed with FNAL on the developing "high-Cp" wires, as well as developing ZrO2 "APC" $Nb_3Sn$. FNAL and LBNL have ongoing SBIRs with Composite Technology Development Inc. related to radiation-tolerant, thermally conductive resin systems. Finally, an excellent example of successful partnership with industry relates to SBIRs that were focused on developing Bi-2212 powders as precursors for powder-in-tube Bi-2212 wires. Two companies, Engi-Mat and MetaMateria, were both able to produce high quality powders; the powder from Engi-Mat is now commercially available and serves as the basis for commercial Bi-2212 wire used by the US MDP. [20]

Many other examples of SBIR collaborations aligned with the US MDP exist. Furthermore, we note that the MDP collaborating institutions have strong, ongoing collaborations with industry not only in the accelerator space but also in synergistic areas like the startup fusion industry. Both LBNL and BNL have grants in programs such as INFUSE and ARPA-E that encourage public-private partnerships in the fusion space. This area has also made use of the amazing technologies developed as part of MDP and synergistic SBIRs such as acoustic and other sensing technologies and also again the common coil test platform that will be used in studies with Commonwealth Fusion. In addition, due to the strong technology development that both LBNL and BNL have carried out for years in the superconducting space, both groups are currently collaborating with GE in the superconducting wind generator space. As GE ventures into the very large coil space, they are utilizing the instrumentation and designs for quench protection that LBNL and BNL have been developing for use in accelerator magnets and other HEP applications like axion magnets.

## Summary

This *2020 Updated Roadmaps for the US Magnet Development Program* will serve to aggressively and fruitfully guide research into high field magnet and superconductor technology for the DOE Office of High Energy Physics for the next 3-4 years. The roadmaps build on the strengths of the MDP team, the experience they have gained over the last four years, and the insights the team has on the critical directions and developments required to advance our field effectively. It further builds on the existing infrastructure and facilities of the MDP, and anticipates efficient expansion of those facilities where needed to deliver on the roadmaps.

The roadmaps also provide vision for high field magnet research in terms of opportunities and challenges, as HEP embarks on the Snowmass community planning process, and taking into consideration the European Strategy for Particle Physics. Finally, we note that the roadmaps provide context for synergistic activities within the DOE Office of Science and with Industry; fusion, in particular, can benefit from, and partner with, these efforts to advance magnet technology so as to enable new and exciting capabilities.

# Bibliography

[1] X. Xu, X. Peng, J. Lee, J. Rochester and M. D. Sumption, "High Critical Current Density in Internally-oxidized Nb3Sn Superconductors and its Origin," *Scr. Mater.,* vol. 186, pp. 317-320, 2020.

[2] S. Balachandran, C. Tarantini, P. J. Lee, F. Kametani, Y. Su, B. Walker, W. L. Starch and D. C. Larbalestier, "Beneficial influence of Hf and Zr additions to Nb4at%Ta on the vortex pinning of Nb3Sn with and without an O source," *Superconductor Science and Technology,* vol. 32, no. 4, 2019.

[3] A. V. Zlobin and D. Schoerling, Nb3Sn Accelerator Magnets - Designs, Technologies, and Performance, Springer, 2019.

[4] A. V. Zlobin and et al, "Development and First Test of the 15T Nb3Sn Dipole Demonstrator MDPCT1," *IEEE Transactions on Applied Superconductivity,* vol. 30, no. 4, 2020.

[5] M. Juchno and et al, "Mechanical Utility Structure for Testing High Field Superconducting Dipole Magnets," *IEEE Transactions on Applied Superconductivity,* vol. 29, no. 5, 2019.

[6] A. V. Zlobin, V. V. Kashikhin and I. Novitski, "Large-aperture high-field Nb3Sn dipole magnets," *Proceedings of IPAC,* 2018.

[7] A. V. Zlobin, J. Carmichael, V. V. kashikhin and I. Novitski, "Conceptual design of a 17 T Nb3Sn accelerator dipole magnet," *Proceedings of IPAC,* 2018.

[8] J. Jiang, G. Bradford, S. Hossain, M. Brown, J. Cooper, E. Miller, Y. Huang, H. Miao, J. Parrell, M. White and e. al., "High-performance Bi-2212 round wires made with recent powders," *IEEE Trans. Appl. Supercond,* vol. 29, 2019.

[9] L. Fajardo, L. Brouwer, S. Caspi, A. Hafalia, C. Hernikl, S. Prestemon, T. Shen, E. Bosque, C. English and R. Hafalia, "Fabrication of Bi-2212 Canted-Cosine-Theta Dipole Prototypes," *IEEE Transactions on Applied Superconductivity,* vol. 29, 2019.

[10] T. Shen and L. Garcia Fajardo, "Superconducting Accelerator Magnets Based on High-Temperature Superconducting Bi-2212 Round Wires," *Instruments,* vol. 4, no. 2, 2020.

[11] A. V. Zlobin, I. Novitski and E. Barzi, "Conceptual Design of an HTS Dipole Insert Based on Bi2212 Rutherford Cable," *Instruments,* vol. 4, no. 2, 2020.

[12] X. Wang, S. A. Gourlay and S. O. Prestemon, "Dipole Magnets Above 20 Tesla: Research Needs for a Path via High-Temperature Superconducting REBCO Conductors," *Instruments,* vol. 3, no. 4, 2019.

[13] V. V. Kashikhin, V. Lombardo and G. Velev, "Magnet Design Optimization for Future Hadron Colliders," *Proceedings of IPAC,* 2019.

[14] L. Brouwer, D. Arbelaez, B. Auchmann, L. Bortot and E. Stubberud, "User defined elements in ANSYS for 2D multiphysics modeling of superconducting magnets," *Superconductor Science and Technology,* 2019.

[15] A. V. Zlobin and et al, "Development and Study of Nb3Sn Wires With High Specific Heat," *IEEE Transactions on Applied Superconductivity,* vol. 29, no. 5, 2019.

[16] R. M. Scanlan and D. R. Dietderich, "Progress and plans for the US HEP conductor development program," *IEEE transactions on applied superconductivity,* vol. 13, no. 2, 2003.

[17] L. D. Cooley, A. K. Ghosh, D. R. Dietderich and I. Pong, "Conductor specification and validation for high-luminosity LHC quadrupole magnets," *IEEE Transactions on Applied Superconductivity,* vol. 27, no. 4, 2017.

[18] S. Hahn, K. Kim, K. Kim, X. Hu and T. Painter, "45.5-tesla direct-current magnetic field generated with a high-temperature superconducting magnet," *Nature,* 2019.

[19] B. N. Sorbom and et al., "ARC: A compact, high-field, fusion nuclear science facility and demonstration power plant with demountable magnets," *Fusion Engineering and Design,* 2015.

[20] J. S. Murtomaki, J. v. Nugteren, A. Stenvall, G. Kirby and L. Rossi, "3-D Mechanical Modeling of 20 T HTS Clover Leaf End Coils—Good Practices and Lessons Learned," *IEEE Transactions on Applied Superconductivity,* vol. 29, no. 5, 2019.

[21] T. Shen, E. Bosque, D. Davis, J. Jiang, M. White, K. Zhang, H. Higley, M. Turqueti, Y. Huang, H. Miao and e. al., "Stable, predictable and training-free operation of superconducting Bi-2212 Rutherford cable racetrack coils at the wire current density of 1000 A/mm2," *Scientific Reports,* vol. 9, 2019.

[22] K. Zhang, H. Higley, L. Ye, S. Gourlay, S. Prestemon, T. Shen, E. Bosque, C. English, J. Jiang, Y. Kim and e. al., "Tripled critical current in racetrack coils made of Bi-2212 Rutherford cables with overpressure processing and leakage control.," *Superconducting Science and Technology,* vol. 31, 2018.

[23] L. Fajardo, L. Brouwer, S. Caspi, S. Gourlay, S. Prestemon, T. Shen and S. Gourlay, "Designs and Prospects of Bi-2212 Canted-Cosine-Theta Magnets to Increase the Magnetic Field of Accelerator Dipoles Beyond 15 T.," *IEEE Transactions on Applied Superconductivity,* vol. 28, 2018.

[24] S. Caspi and et al., "Canted-Cosine-Theta Magnet (CCT) A Concept for High Field Accelerator Magnets," *IEEE Transactions on Applied Superconductvity,* vol. 24, 2014.

[25] N. Diachenko, "Stress management in high field dipoles," *Proceedings, Particle Accelerator Conference, Vancouver,* 1997.

# Appendices

## Appendix I: Milestone tables

The following milestone tables, along with Figure 12, provide tangible deliverables with target dates for completion, and are compatible with the *Updated Roadmaps* in Figure 9. The milestones in bold serve as major targets for the various areas.

*Table 2. Milestones for the Stress-Managed Cosine-Theta (SMCT) effort within the $Nb_3Sn$ area of the MDP.*

| Milestone # | Description | Target |
|---|---|---|
| **AI-M1a** | Development and test of stress management concept using a 2-layer large-aperture and 4-layer small-aperture cos-theta coils and dipole mirror structure | March 2022 |
| **AI-M2a** | Development, fabrication and test of stress management concept in a 2-layer 120-mm dipole with the field up to 11 T. | April 2023 |
| AI-M3a | Assembly and test of stress-management concept in a 4-layer 60-mm 17 T dipole with stress management. | April 2024 |

*Table 3. Milestones for the Canted-Cosine-Theta (CCT) effort within the $Nb_3Sn$ area of the MDP.*

| Milestone # | Description | Target |
|---|---|---|
| AI-M1b | Improve numerical modeling of magnet mechanics to accurately predict stresses when surface failure/delamination can occur. | December 2020 |
| **AI-M2b** | Use CCT subscale program to increase understanding and improve training in stress managed magnets. | July 2021 |
| AI-M3b | Design of CCT6 4-layer magnet with target field of 13 T and bore diameter of 120 mm (with feedback from desired diameter from HTS components of MDP). Design of CCT6 will profit from feedback provided by subscale testing and improved modeling methods. | January 2022 |
| **AI-M4b** | Fabrication and test of CCT6 4-layer CCT magnet. | June 2023 |

*Table 4. Milestones for the CCT effort within the Bi-2212 area of the MDP.*

| Milestone # | Description | Target |
|---|---|---|
| **AIIa-M1a** | Build and test two 2.4 T, 40 cm long BIN5c dipole magnets. | Dec 2020 and March 2021 |
| AIIa-M2a | Hybrid magnet test with a total field generation of 8-10 T at 4.2 K. Assemble and test BIN5c in the background field of the $Nb_3Sn$ CCT5 | Dec 2021 |
| **AIIa-M3a** | Build and test two 3.5 T, 80 cm long, Bi-2212 dipole magnet with 17-strand, 7.8 mm wide Rutherford cables. | July 2021 and Jan 2022 |
| **AIIa-M4a** | Build and test two 5 T, 80 cm long, Bi-2212 dipole magnet with 27-strand, 12 mm wide Rutherford cables. | July 2022 and Jan 2023 |
| **AIIa-M5a** | Hybrid magnet test with a total field generation of >14.5 T at 4.2 K. Assemble and test magnets from AI-M3c and M4c inside a background field of the 120 mm, 11 T $Nb_3Sn$ magnet from the area I. | Sept 2023 |

*Table 5. Milestones for the Stress-Managed Cosine-Theta (SMCT) effort within the Bi-2212 area of the MDP.*

| Milestone # | Description | Target |
|---|---|---|
| AIIa-M1b | Study strand damages due to cabling, transverse pressure dependence | April 2022 |
| AIIa-M2b | Fabricate the first 2-layer 17-mm aperture Bi-2212 coil using LBNL cable. Coil test independently and inside a 60-mm aperture 2-layer $Nb_3Sn$ dipole coil in mirror configuration. | July 2022 |
| **AIIa-M3b** | Fabricate the 2nd 2-layer 17-mm aperture Bi-2212 coil using optimized Bi-2212 cable, coil structure, materials and technologies. Coil test independently and inside a 60-mm aperture 4-layer $Nb_3Sn$ dipole coil in mirror configuration. | December 2022 |
| **AIIa-M4b** | Fabricate another 2-layer Bi-2212 coil using optimized Bi-2212 cable and coil structure. Bi-2212 coil test independently and inside a 60-mm aperture 4-layer Nb3Sn dipole coil. | September 2024 |

*Table 6. Milestones for OPHT facility upgrade and Rutherford-cable-based solenoid development.*

| Milestone # | Description | Target |
|---|---|---|
| **AIIa-M1c** | Renegade OPHT facility upgrade. 1 m and 250 mm hot zone. | Dec 2020 |
| AIIa-M2c | 16 T Rutherford cable based solenoid development | Dec 2020 |
| AIIa-M3c | 20 T Rutherford cable based solenoid development | June 2021 |
| AIIa-M4c | 25 T Rutherford cable based solenoid development | June 2022 |

*Table 5. Milestones for the REBCO effort within the HTS area of the MDP.*

| Milestone # | Description | Target |
|---|---|---|
| | **Magnet technology development** | |
| **AIIb-M1** | Test of CORC® subscale common coil in 10 T background field | June 2021 |
| AIIb-M2 | Demonstrate first COMB (Conductor on Molded Barrel) technology | September 2021 |
| **AIIb-M3** | CORC® CCT to reach 5 T dipole field | December 2021 |
| AIIb-M4 | Complete design study of a 8 T REBCO dipole magnet | December 2021 |
| **AIIb-M5** | Complete COMB insert test | May 2022 |
| AIIb-M6 | Generate 1 T with REBCO insert in a background field of 8 T from $Nb_3Sn$ CCT5 | June 2022 |
| **AIIb-M7** | COMB performance demonstration | March 2023 |
| AIIb-M8 | REBCO magnet generate a 8 T dipole field stand-alone | March 2023 |
| | **Conductor characterization** | |
| AIIb-M9 | CORC® wire quench study at BNL 10 T common coil magnet | December 2020 |
| AIIb-M10 | Impact of Lorentz load on CORC® wires using ASC 14 T solenoid | June 2021 |
| | **Key assumptions: infrastructure availability** | |
| **AIIb-M11** | Commission $Nb_3Sn$ CCT5 test platform | June 2021 |
| **AIIb-M12** | 120 mm aperture 10 – 12 T $Nb_3Sn$ magnet | June 2022 |
| **AIIb-M13** | Hybrid test platform with outsert magnet available at FNAL | December 2022 |

*Table 6. Milestones for the 20 T hybrid magnet design and comparative analysis effort within the Technology area of the MDP.*

| Milestone # | Description | Target |
|---|---|---|
| AIIIa-M1 | Review previous work on hybrid magnets, and identify the design and comparison criteria. In addition, perform preliminary designs through analytical models. | July 2021 |
| **AIIIa-M2** | Perform 2D finite element models to study the different design options, focusing on magnetic, mechanical and quench protection studies | November 2021 |
| **AIIIa-M3** | Comparison analysis, including considerations about fabrication, cost, integration and testing | January 2023 |

*Table 7. Milestones for the Advanced Modeling effort within the Technology area of the MDP.*

| Milestone # | Description | Target |
|---|---|---|
| AIIIb-M1 | Identification and/or development of advanced finite element techniques for mechanical interfaces (such as cohesive zone elements) | May 2021 |
| AIIIb-M2 | Extension of existing codes to model HTS quench behavior and magnetization as required for hybrid magnet protection studies | May 2021 |
| AIIIb-M3 | Integration of advanced contact elements in existing CCT and SMCT mechanical models | June 2021 |
| AIIIb-M4 | Benchmarking of new codes with existing hybrid magnet test data | Sept 2021 |
| AIIIb-M5 | Report correlating interface modeling results with experimental magnet training and summarizing desired interface conditions for stress managed designs | December 2021 |
| **AIIIb-M6** | Design and optimization of quench protection systems for planned hybrid tests in the $Nb_3Sn$ and HTS roadmaps | May 2022 |
| AIIIb-M7 | Efficient parallelization of new and existing HTS codes to leverage DOE high-performance computing resources | December 2022 |
| AIIIb-M8 | Application of multi-scale, mechanical strain studies to refine strain-based conductor limitations from the strand to magnet level for both $Nb_3Sn$ and HTS designs | December 2022 |
| AIIIb-M9 | Development of new methods for modeling of HTS which leverage advanced numerical techniques such as adaptive meshing and unfitted finite element methods (such as XFEM) | March 2023 |
| **AIIIb-M10** | Creation of a hierarchical, multiphysics toolbox for HTS cable modeling with increasing levels of complexity and computational requirements (from equivalent circuit models to finite element techniques) | March 2023 |

*Table 8. Near term milestones for the Magnet Materials section. The focus initially is on novel impregnation resins and on interface treatments.*

| Milestone # | Description | Target |
|---|---|---|
| **AIIIc-M1** | Demonstration of High Viscosity (Thermoplastic) Resin Systems for Superconducting Magnets | June 2021 |
| AIIIc -M2 | Down-selection and Fabrication of CCT based on the above | November 2021 |
| AIIIc -M3 | Evaluation of surface treatments and interface modification | December 2021 |
| **AIIIc -M4** | Test of a CCT implementing interface modifications | July 2022 |

*Table 9. Near term milestones for the Novel Diagnostics section.*

| Milestone # | Description | Target |
|---|---|---|
| AIIId-M1 | Development of a new generation of self-calibrating acoustic emission diagnostics hardware | December 2020 |
| AIIId-M2 | Finalizing software algorithms for acoustic data analysis, completing analysis for the CCTs and 15T dipole | December 2020 |
| AIIId-M3 | Development and test of a linear quench localization sensor on a Bi-2212 subscale and/or ReBCO CCT series | March 2021 |
| **AIIId-M4** | Test of a large-scale Hall array and imaging current distribution in HTS tape stacks and coils | May 2021 |
| AIIId-M5 | Completing spot heater studies to improve voltage-based diagnostics and address “silent” quenches | July 2021 |
| AIIId-M6 | Demonstration of inverse acoustics-based probing of interfaces in a dedicated small-scale coil | September 2021 |
| AIIId-M7 | Development of multi-element and flexible quench antennas and localization of quenches in using flexible quench antenna arrays | September 2021 |
| AIIId-M8 | Characterization of training-like behavior in different impregnation materials under load using a Transverse Pressure Insert (TPI) measurement system | December 2021 |
| AIIId-M9 | Development and test of a standalone acoustic quench detection and localization FPGA-based system | December 2021 |
| **AIIId-M10** | Development and test of a non-rotating new magnetic probe prototype | December 2021 |
| **AIIId-M11** | Demonstration of a programmable fully-cryogenic FPGA “smart” sensor core with digital readout and analog front-end (SQUID) amplifiers | December 2021 |
| **AIIId-M12** | Calibration of FBG fibers in a small cryostat. Installation on an MDP magnet and strain measurement during a quench. Design a proof of principle experiment for quench 3D spatial detection and coil azimuthal strain mapping and install fiber on MDP magnet. Use fibers for energy spectrum analysis and HTS quench detection. | December 2022 |

*Table 10. Near term milestones for the Training Reduction subsection within the Technology area..*

| Milestone # | Description | Target |
|---|---|---|
| AIIIe-M1 | Commissioning of QCD | May 2021 |
| AIIIe-M2 | First Ultrasound based test | May 2021 |
| AIIIe-M3 | First high-Cp cable fabrication | September 2021 |
| AIIIe-M4 | First magnet test with QCD | September 2021 |
| **AIIIe-M5** | Results from High-Cp cable studies | December 2021 |
| AIIIe-M6 | Optimized strand and cable FEM simulations | December 2021 |
| **AIIIe-M7** | First CCT test with QCD | February 2022 |
| AIIIe-M8 | High-Cp wire and tape optimized versions | May 2022 |
| AIIIe-M9 | Fabrication of first coil with High-Cp conductor | September 2022 |
| AIIIe-M10 | Design of a dedicated device/technique using vibrational methods | September 2022 |
| AIIIe-M11 | Design of a "cable/stack" testing device and samples | January 2023 |
| AIIIe-M12 | QCD preparations and test on a large magnet | February 2023 |
| **AIIIe-M13** | Fabrication of a "cable/stack" testing device | September 2023 |

*Table 11 Near term milestones for the development of Bi-2212 wires.*

| Milestone # | Description | Target |
|---|---|---|
| CPRD-Bi-2212-1 | Develop production-like specifications, quality-control methods, and quality assurance. QC methods may need assistance of University programs. | 2021 |
| CPRD-Bi-2212-2 | Achieve consistent longer pieces, stable critical current, and consistent overall production in larger billets | 2023 20 kg billets<br>2025 50 kg billets |
| CPRD-Bi-2212-3 | Facilitate innovations and improvements | As needed |

*Table 12 Near term milestones for the development of advanced $Nb_3Sn$ wires.*

| Milestone # | Description | Target |
|---|---|---|
| CPRD-$Nb_3Sn$-1 | Transfer of LDRD, ECRA, and University research ideas to development scale proposals at major industries. | 2021(better 2022?), contingent on alloy availability |
| CPRD-$Nb_3Sn$-2 | Scale successful industry development to production level | 2023 half billets<br>2025 full billets |
| CPRD-$Nb_3Sn$-3 | Facilitate combined approaches at development level | 2022 and beyond |

*Table 13 Near term milestones for the development of REBCO wires and cables.*

| Milestone # | Description | Target |
|---|---|---|
| CPRD-REBCO-1 | Address coated conductor and cable aspects that improve flexibility and bend tolerance in CCT magnets, e.g. by addition of a lubricant with low inter-conductor resistance | 2021 |
| CPRD-REBCO-2 | Achieve consistent cross-section and longer pieces, in the present thin, narrow configuration, to facilitate longer cables | 2022 |
| CPRD-REBCO-3 | Adapt pinning additives to needs at 4 K and high field in isotropic cables, such as by addition of Y2O3 nanodots instead of BaZrO3 nanorods. | 2023 |

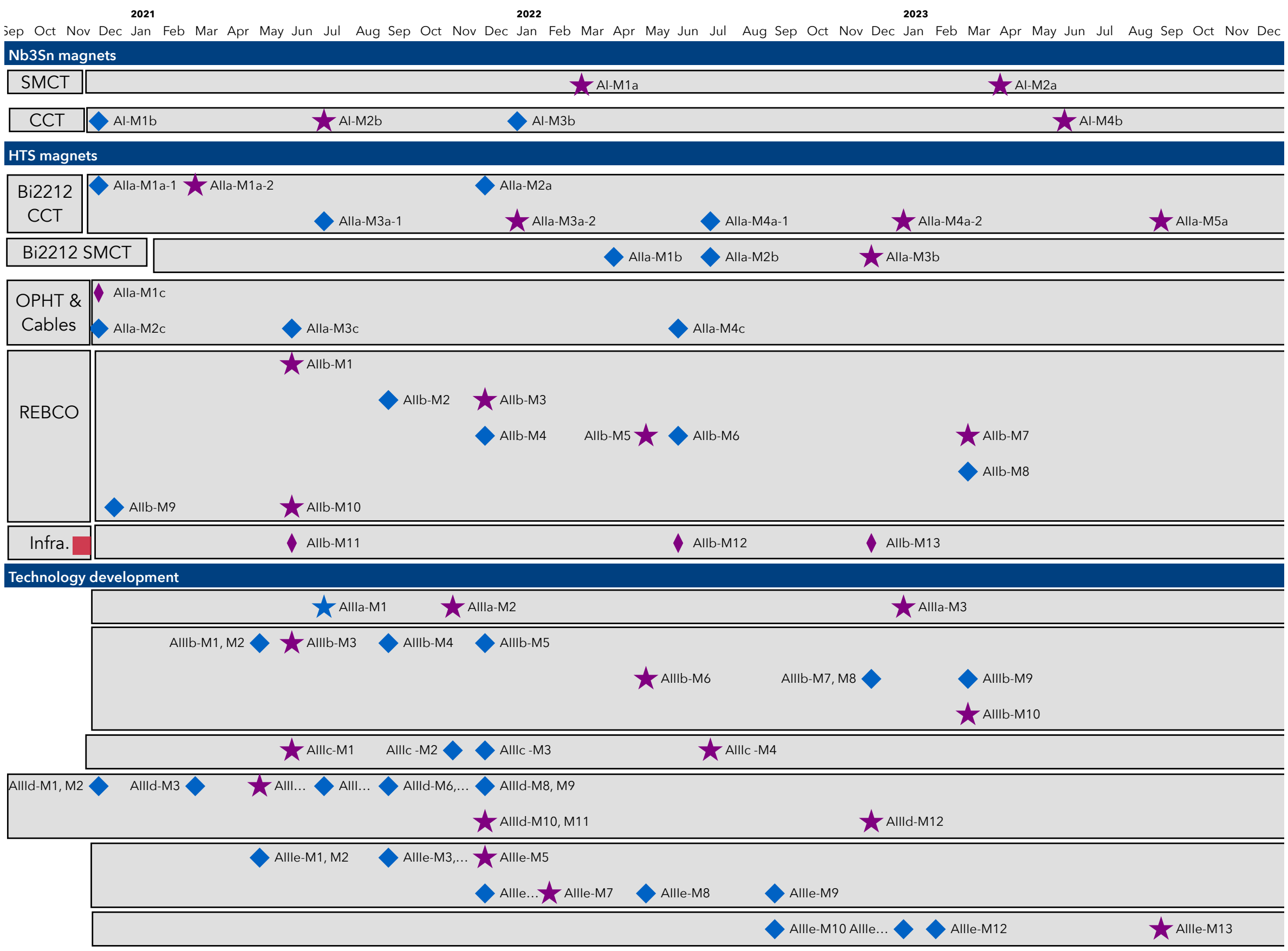

*Figure 12 Milestone timeline for 2021-2023. Red stars identify major milestones; red triangles identify infrastructure milestones.*

## Appendix II: The MDP Community

The MDP is composed of Scientists, Engineers, and Technical staff from the participating laboratories and universities. Due to the nature of the program, staff typically contribute a fraction of their time to research for the US MDP program, with the complement applied to other research and/or project activities at their respective institutions. A picture of members who attended the 2020 US MDP Collaboration meeting in Berkeley in Feb. 2020 is shown in Figure 13. Note that not all MDP participants from collaborating institutions could attend; nevertheless, the collaboration meeting had more than 50 participants, with ~36 individual presentations as well as a poster session to enable further technical discussions.

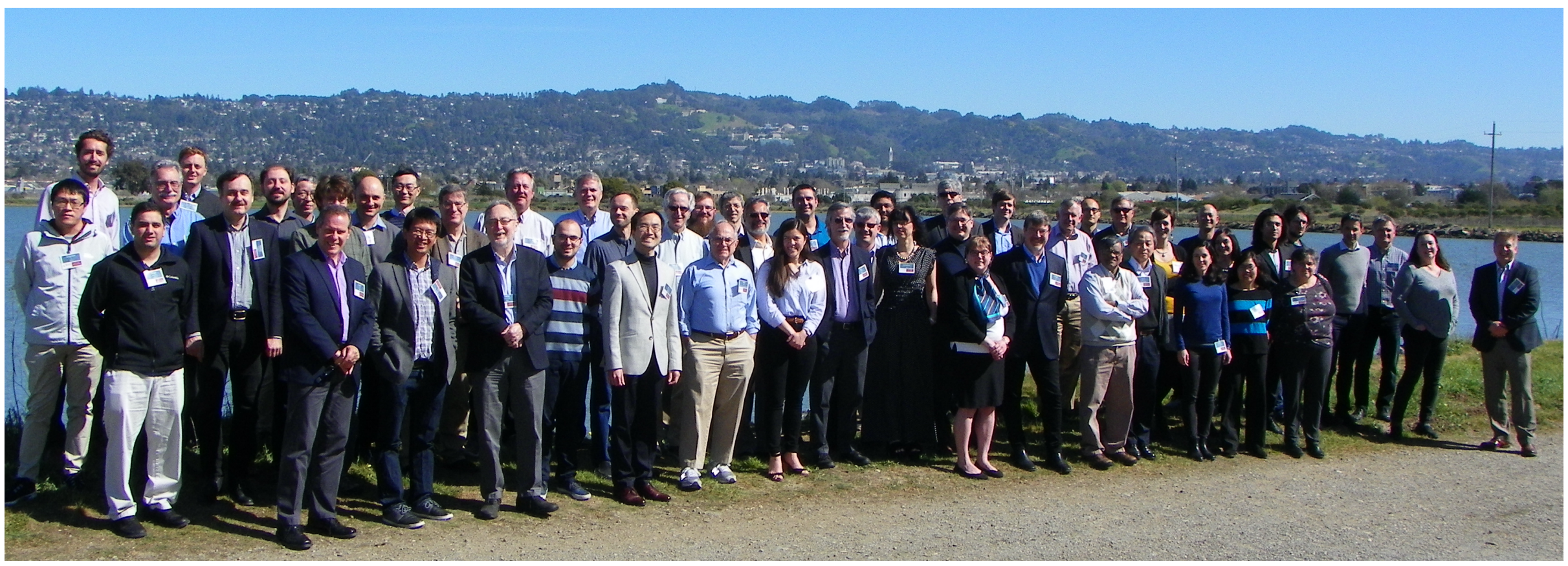

*Figure 13. Photograph of the 2020 US MDP Collaboration meeting held in February 2020, at the Doubletree Hotel, Berkeley Marina, Berkeley, California.*

## Appendix III: Management Structure

The US MDP management structure is designed to integrate the expertise and facilities from the collaborating institutions to rapidly develop high field accelerator magnet technology for the DOE Office of High Energy Physics (see Figure 14). The program is led by LBNL, with partner DOE Laboratories FNAL and BNL, as well as the University program from ASC/NHMFL.

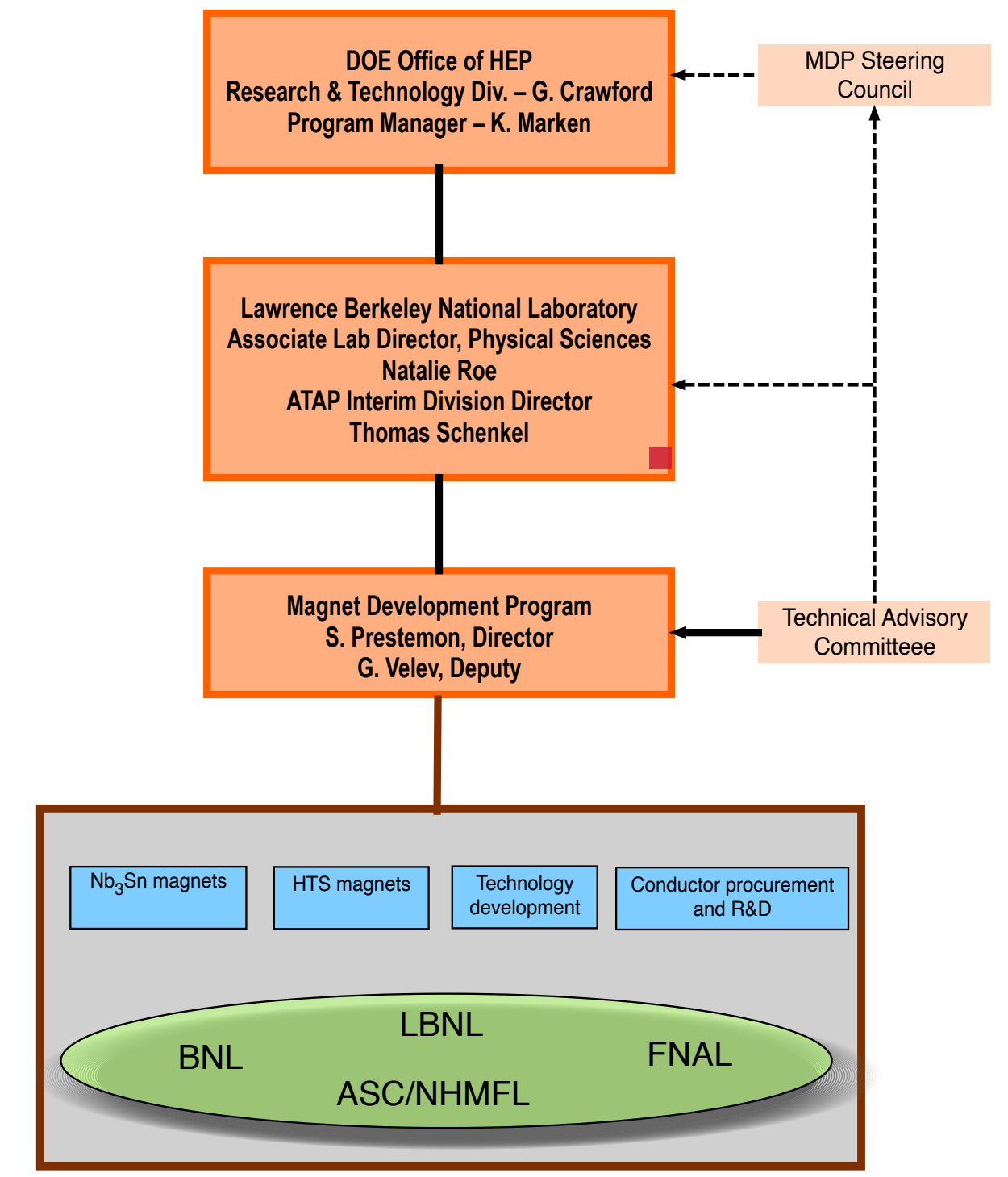

*Figure 14. Management structure for the US MDP.*

Internal to the program, a management group of seven senior scientists, composed of members from all collaborating institutions, convenes on a weekly basis to review progress and plans for the program, and to identify issues and opportunities for consideration.

At the lead-laboratory level, the MDP Director reports to the Director of the host LBNL Division, the Accelerator Technologies and Applied Physics Division, and the Associate Laboratory Director for Physical Sciences. The Division Director and Associate Director, in turn, report to the MDP Program Manager in the Office of High Energy Physics.

To provide technical and strategic guidance and oversight, two committees have been established:

I. A Technical Advisory Committee (TAC; see Table 1), composed of technical experts in the field and reporting to the MDP Director, provides guidance on program planning, technical progress, and program strategy. The TAC meets with the MDP management team 2-3 times a year, and attends the yearly MDP Collaboration Meeting.
II. A Steering Council, reporting to DOE-OHEP, is composed of Laboratory Directors (or a delegate) and two external members chosen by DOE-OHEP (see Table 15). The Chair of the Steering Council is selected by DOE-OHEP. The steering council, which hears from the TAC as well as from the MDP Director on a yearly basis, provides guidance on overall program strategy, inter-laboratory integration, and national and international collaborations.

In addition, as described above in the section *Area IV: Conductor Procurement and Research & Development (CPRD)*, an Advisory Committee (see Table 16) provides guidance on conductor research directions for the MDP; the committee reports to the MDP Director.

*Table 14. Membership of the Technical Advisory Committee.*

| | |
|---|---|
| Andrew Lankford (Chair) | University of California, Irvine |
| Giorgio Apollinari | Fermi National Accelerator Laboratory |
| Joseph Minervini | Massachusetts Institute of Tech. (retired) |
| Mark Palmer | Brookhaven National Laboratory |
| Davide Tommasini | CERN |
| Akira Yamamoto | KEK & CERN |

*Table 15. Membership of the Steering Council.*

| | |
|---|---|
| Harry Weerts (Chair; DOE representative) | Argonne National Laboratory (retired) |
| Tor Raubenheimer (DOE representative) | SLAC National Accelerator Laboratory |
| Michael Witherall (or designee) | Lawrence Berkeley National Laboratory |
| Nigel Lockyer (or designee) | Fermi National Accelerator Laboratory |
| Gregory Boebinger (or designee) | Florida State University / NHMFL |
| Doon Gibbs (or designee) | Brookhaven National Laboratory |

*Table 16. Membership of the Conductor Procurement and R&D (CPRD) Advisory Panel.*

| | |
|---|---|
| Lance Cooley (Head) | Applied Superconductivity Center/NHMFL |
| Ian Pong (Secretary & Record-keeping) | Lawrence Berkeley National Laboratory |
| David Larbalestier | Applied Superconductivity Center/NHMFL |
| Matt Jewell | University of Wisconsin – Eau Claire |
| Vlad Matias | I-Beam Materials |
| Vito Lombardo | Fermi National Accelerator Laboratory |